\documentclass{epl}
\usepackage{epsf}
\newcommand{\sbf}[1]{\mbox{{\scriptsize$\bf{#1}$}}}
 \def\div{\rm d\hspace{-0pt}i\hspace{-.5pt}v}
 \def\inter{\rm i\hspace{1pt}nt}

\def\gl{\displaystyle{\raisebox{-.2em}{$>$}\atop\raisebox{.2em}{$<$}}}
\def\gl{\raisebox{-2.2mm}{~\hspace{-2.8mm}
\raisebox{2mm}{$\begin{array}{c}\scriptsize>
\\[-2.6mm]\scriptsize<\end{array}$}\hspace{-1.6mm}}}
\def\qn{x}
\def\phin{q}
\def\dDelta{\,\dot{}\Delta\,}
\def\dDeltad{\,\dot{}\Delta\dot{}\,}
\def\Deltad{\,\Delta\dot{}\,}
\def\ddDelta{\,\ddot{}\Delta\,}
\def\Deltadd{\,\Delta\ddot{}\,}
\def\ddDeltad{\,\ddot{}\Delta \dot{}\,}
\def\Deltaddd{\,\Delta\ddot{} \dot{} \,}
\def\ve{\frac{\varepsilon^2}{2} }

\def\comment#1{}

\title{Integrals over
Products of Distributions \\
from Manifest Coordinate Invariance of\\
Perturbation
Expansions of Path Integrals in Curved Space}
\shorttitle{Integrals over
Products of Distributions ...}
\author{H.~Kleinert\thanks{E-mail: kleinert@physik.fu-berlin.de}\inst{1} and
     A.~Chervyakov\thanks{On leave from LCTA, JINR, Dubna, Russia,
                   E-mail: chervyak@physik.fu-berlin.de}\inst{1}}
                     \institute{ Freie Universit\"at Berlin\\
          Institut f\"ur Theoretische Physik\\
          Arnimallee 14, D-14195 Berlin}
\pacs{3.65.-w}{Quantum mechanics}%{First pacs description}
\pacs{2.90.+p}{Other topics in mathematical methods in physics}%{Second pacs
\pacs{2.30.Qy}{Integral transforms and operational calculus}%{Third pacs
\begin{document}

\maketitle

\begin{abstract}
We show that the requirement
of manifest coordinate invariance
of perturbatively defined quantum-mechanical  path integrals
in curved space leads to
an extension of the
theory  of  distributions
by  specifying unique rules for integrating products
of distributions.
The rules are derived
using equations of motion and
partial integration, while
keeping track of
certain minimal features
stemming
from the
unique definition of all singular integrals
 in $1 - \epsilon $
dimensions.
Our rules guarantee complete agreement with
 much more cumbersome calculations
  in $1- \epsilon  $
dimensions where the limit $ \epsilon \rightarrow 0$
is taken at the end.
In contrast to
our previous papers
where we solved the same
problem
for an infinite time interval or
zero temperature,
we consider here
the more involved case of
finite-time (or non-zero temperature) amplitudes.
\end{abstract}

\section{Introduction}

Until recently,
a coordinate-independent
definition
of quantum mechanical path integrals in curved space
existed
only in the time-sliced
formulation  \cite{PI}.
This is in contrast to field-theoretic
path integrals between two and four spacetime dimensions
which are well-defined
in continuous spacetimes
by perturbation expansions.
Initial
difficulties in guaranteeing coordinate independence
were  solved
by
't~Hooft and Veltman \cite{4}
using
dimensional regularization
with minimal subtractions (for a detailed
description of this method see the textbook \cite{V}).
Coordinate  independence
 emerges
after calculating all Feynman integrals
in an arbitrary
number of dimensions $d$,
and continuing
the results to the $d=4$.
Infinities occuring in the limit are absorbed into parameters of the action.

In contrast, and surprisingly,
numerous attempts
\cite{5,6,7,8,9,10,11,12,13}
to define the simpler  {\em quantum mechanical\/}
path integrals  in curved space
by perturbation
expansions
encountered problems
in evaluating
the Feynman integrals. Although all physical properties
are
finite and uniquely determined by Schr\"odinger theory,
the  Feynman integrals in the expansions are highly
singular and mathematically undefined.
When evaluated in momentum space, they yield different results depending on the order of integration.
Various definitions
 chosen by earlier authors \cite{5,6,7,8,9,10,11,12,13}
were not manifestly coordinate-independent,
and this could only be cured
by
 adding
 coordinate-dependent
``correction terms"
to the classical action---a highly unsatisfactory procedure
violating the
basic Feynman rule
that physical amplitudes should consist
of a sum over all paths of
phase factors
$e^{i{\cal A}}$
whose exponents
contains only
the
coordinate-independent
classical action along the paths.

The first
satisfactory perturbative  definition
of path integrals in curved space
was found  only recently
by us \cite{2,3,3a}.
The results enabled us
 to
set up
simple  rules for
treating integrals over products of distributions
in one dimension
to ensure
coordinate invariance \cite{3a}.
These rules
were
derived for
path integrals on an infinite time interval
or zero temperature, where we could apply
 most directly the dimensionally continued integration rules
of 't~Hooft and Veltman \cite{4} in momentum space.

In a recent paper~\cite{14},
the authors
of
\cite{8}
and \cite{12}
have adapted the methods
developed in our first two papers
\cite{2,3}
to
the calculation of finite-time amplitudes
(see also \cite{15,BD}).
In doing this they have, however,
not taken advantage of the
great simplifications
brought about by the
developments in our third paper \cite{3a}
which make  explicit evaluations of Feynman integrals
in $d=1- \epsilon $ dimensions superfluous.
The purpose of the present work
is to
show how this happens,
Thus we shall derive
 rules
for calculating
 integrals over products
of distributions
which automatically guarantee coordinate independence.
All integrals will be evaluated
in one dimension, after
having been brought to
a regular form by some
trivial manipulations
which require only a small residual information
on the
initial $1- \epsilon $ -dimensional nature of the
Feynman integrals.

Consider the short-time amplitude
of a particle in curvilinear coordinates
\begin{equation}
(q_b\tau _b|q_a\tau _a) = \int  {\cal D} \phin  (\tau) \sqrt{g} \,
e^{-{\cal A} [\phin ]},
\label{wbn@@1}\end{equation}
where ${\cal A} [\phin ]$ is the euclidean action
of the form
\begin{equation}
{\cal A} [\phin ]=\int_{\tau _a}^{\tau _b} d\tau \left[
\frac{1}{2}g_{ij}(q(\tau ))
\dot q^{i} (\tau )
\dot q^{j} (\tau )+V(q(\tau ))\right].
\label{wbn@@}\end{equation}
The dots denote $\tau $-derivatives, $g_{ij}(q)$ is
a metric, and $g=\det g$
its determinant.
The path integral may formally
be defined perturbatively as follows:
The metric
$g_{ij}(q)$
and the potential $V(q)$
are expanded
around some point $q_{0}^{i}$
near $q_a$ and $q_b$
in powers of $ \delta q^i \equiv q^i - q_{0}^{i}$.
After this, the action ${\cal A} [\phin ]$
is separated into a free  part
$
{\cal A}_0[q_0; \delta \phin ]\equiv(1/2)\int_{\tau _a}^{\tau _b} d\tau\,
g_{ij}(q_0)
\dot q^{i} \dot q^{j}
$,
and an interacting part
${\cal A}_{\inter}[q_0; \delta \phin ]\equiv
{\cal A}[ \phin ]-
{\cal A}_0[q_0; \delta \phin ]$.

A simply curable ultraviolet (UV)
 divergence problem is encountered
in the square root in the
measure of
functional integration in
(\ref{wbn@@1}).  Taking it into the exponent and expanding
in powers of $ \delta q$, it  corresponds to
an effective action
 \begin{eqnarray}
 {\cal A}_{ \sqrt{g} }=-\frac{1}{2} \delta (0)
\int_{\tau _a}^{\tau _b} d\tau \log [ g(q_0+ \delta q)/g(q_0)],
\label{wbn@@acteff}\end{eqnarray}
which contains the
$ \delta $-function at the origin
$ \delta (0)$.
This infinite quantity
represents formally
the inverse infinitesimal lattice spacing on the time axis,
and is
equal to the
momentum
integral
$\delta (0)\equiv \int d p/(2\pi)$.
 With (\ref{wbn@@acteff}),
the path integral  (\ref{wbn@@1})
takes the form
\begin{equation}
(q_b\tau _b|q_a\tau _a)
 = \int  {\cal D} \phin  (\tau) \,
e^{-{\cal A} [ \phin ]-{\cal A}_{ \sqrt{g} } [ \phin]}
 = \int  {\cal D}\delta  \phin  (\tau) \,
e^{-{\cal A} [q_0+\delta \phin ]-{\cal A}_{ \sqrt{g} } [q_0+\delta \phin]}\,.
\label{wbn@@1a}\end{equation}

The main problem arises in the
expansion of the amplitude in powers of the interaction.
For simplicity, we shall
 set
$\tau _a=0,\,\tau _b= \beta $, as in thermodynamics
and assume $q_b=q_a=q_0=0$.
Performing all Wick contractions,
the origin to origin amplitude
$(0\, \beta |0\,0)$ is expressed as
a sum of loop diagrams.
There are interaction terms involving
$ \delta {\dot q}^2  \delta q^n$
which lead to Feynman integrals
over products of distributions.
The diagrams contain
four types of lines representing
the correlation functions
\begin{eqnarray}
\Delta (\tau ,\tau ')&\equiv& \langle  \delta \phin (\tau ) \delta
\phin (\tau ')\rangle=
\hspace{0mm}\raisebox{-1mm}{\mbox{\input 1.tps }} ,~\label{wbn@p1} ~~~~~~\Delta\dot{}\,(\tau,\tau ')
\equiv \langle  \delta \phin (\tau ) \delta \dot\phin (\tau ')\rangle
=\hspace{0.3mm}\raisebox{0.5mm}{\mbox{\input 3spo.tps }}\! ,
\label{wbn@p2r}
{}~\nonumber \\
\dot{} \Delta\,(\tau,\tau ')&\equiv&\langle \delta  \dot \phin (\tau ) \delta
\phin (\tau ')\rangle
=\hspace{-1pt}\hspace{0mm}\raisebox{-1mm}{\mbox{\input 3o.tps }}\!\! ,~\label{wbn@p2}~~~~~ \hspace{2pt}
 \dot{}\Delta\dot{}\, (\tau,\tau ')\equiv
\langle \delta  \dot \phin (\tau )  \delta \dot\phin (\tau ')\rangle
=\,\hspace{0mm}\raisebox{-1mm}{\mbox{\input 2.tps }}.~\label{wbn@p3}
\label{wbn@prop}\end{eqnarray}
The right-hand sides show the line symbols
to be used in  Feynman diagrams.

The first
correlation function $
  \Delta (\tau ,\tau ') = \Delta (\tau ', \tau  )
$
is determined
by the free part ${\cal A}_0 [q_0;  \delta q]$
of the action. It is the Green function of the
equation of motion
\begin{equation}
\ddot {}\Delta (\tau ,\tau ')=
 \Delta \ddot{}\, (\tau ,\tau ')=- \delta (\tau -\tau '),
\label{wbn@eom}\end{equation}
satisfying the Dirichlet boundary conditions
\begin{equation}
  \Delta (0,\tau ') = \Delta ( \beta , \tau ')=0,~~~~
  \Delta (\tau ,0) = \Delta (\tau , \beta )=0.
%\label{wbn@@}
\end{equation}
Explicitly, it reads
\begin{equation}
 \Delta (\tau ,\tau ')=
 \Delta (\tau ',\tau )=
\frac{1}{2}\left[ - \epsilon (\tau -\tau ')(\tau -\tau ')+
\tau +\tau '\right] -\frac{\tau \tau '}\beta,
%=\Theta(\tau -\tau ')\left(1-\tau / \beta \right)\tau '
%+\Theta(\tau '-\tau )\left(1-\tau '/ \beta\right)\tau,
\label{wbn@p4}\end{equation}
where
$\epsilon (\tau - \tau ')$ is the antisymmetric distribution which
is equal to $\pm 1$ for $\tau\gl\tau '$ and vanishes
 at the origin.

The second and third
correlation functions
$  \dDelta (\tau ,\tau ')$ and $\Deltad (\tau , \tau ' )$ are
\begin{equation}
 \dDelta(\tau ,\tau ') =
     - \frac{1}{2} \epsilon (\tau - \tau ')
+\frac{1}{2}-\frac{\tau '}\beta,
{}~~~
 \Deltad(\tau ,\tau ') =
     \frac{1}{2} \epsilon (\tau - \tau ')
+\frac{1}{2}-\frac{\tau }\beta=
\dDelta(\tau' ,\tau )\,,
\label{wbn@p5}\end{equation}
with a discontinuity at $\tau =\tau '$.
Here and in the following,
dots on the right and left
of $ \Delta (\tau ,\tau ')$ denote
time derivatives with respect to
$\tau $ and $\tau '$, respectively.

The fourth correlation function
$  \dDeltad (\tau ,\tau ')$
is simply
\begin{equation}
 \dDeltad (\tau, \tau ') =
  \delta(\tau -\tau ') -1/ \beta .
\label{wbn@p7}\end{equation}
The
$ \delta$-function
arises from the derivative
$ \delta (\tau - \tau ' )
= \dot\epsilon (\tau - \tau ' )/2$.
Its value at the origin must be equal to the prefactor
$ \delta (0)$ of the effective action (\ref{wbn@@acteff})
of the  measure to cancel
all ultraviolet (UV) infinities.
Note the close similarity  of (\ref{wbn@p7})
to
the equation of motion (\ref{wbn@eom}).

The difficulty in calculating
the loop integrals over products
of such distributions is
best
illustrated by observing the lack
of reparametrization invariance of the path integral
of a free particle in $n$-dimensional curvilinear coordinates
first done
by
Gervais and Jevicki \cite{5}, Salomonson
\cite{6}, and recently also by
Bastianelli, van Nieuwenhuizen, and collaborators
\cite{7,8,9,10,11,12,13}.
The basic ambiguous integrals
causing problems
arise from the
 two-loop diagrams
\begin{eqnarray}
&&\hspace{0mm}\raisebox{-1mm}{\mbox{\input the112.tps }}:~~~~~~
  I_{14} = \int_0^ \beta  \int_0^ \beta  d\tau\, d\tau '
\dDelta (\tau ,\tau ')  \Deltad  (\tau ,\tau ' ) \dDeltad
 (\tau ,\tau '),
\label{wbn@7*}\\
 &&\hspace{0mm}\raisebox{-1mm}{\mbox{\input the022.tps }}:~~~~~~ I_{15} = \int _0^ \beta \int_0^ \beta
d\tau\,d\tau '  \Delta  (\tau ,\tau ') \dDeltad
 ^2 (\tau ,\tau ').
\label{wbn@8*}\end{eqnarray}
It is shown in Appendix A
that
the requirement of coordinate independence
implies that these integrals
have the values
\begin{equation}
I_{14} =  \beta /24,~~~~~
I_{15}^{R} = -  \beta /8,
 \label{wbn@I12}\end{equation}
%
%\cite{5,6,7,8,9,10,11,12,13},
where the superscript $R$ denotes the finite part of
an integral.

Let us demonstrate
that these values are incompatible
with partial integration and the equation
of motion (\ref{wbn@eom}).
In  the integral (\ref{wbn@7*}),
 we use the symmetry $\ddDelta (\tau,\tau ') =
\Deltadd (\tau,\tau ')$,
 apply  partial integration twice
taking care of
nonzero boundary terms,
and obtain
on the one hand
\begin{eqnarray}
\!\hspace{-1cm} I_{14} & = & \frac{1}{2}  \int_0^ \beta
\int_0^ \beta  d\tau\,  d\tau '\, \dDelta
 (\tau ,\tau ') \frac{d}{d\tau } \left[ \Deltad ^2 (\tau
  ,\tau ') \right] % = \nonumber \\
  = - \frac{1}{2} \int_0^\beta  \int_0^\beta  d\tau \, d\tau '  \Deltad ^2
(\tau .\tau ')
         \,\,\ddDelta  (\tau ,\tau ')  \nonumber \\
 \hspace{-1cm}& = &  - \frac{1}{6}  \int_0^\beta  \int_0^\beta  d\tau \, d\tau
' \frac{d}{d\tau '}
     \left[ \Deltad ^3 (\tau ,\tau ') \right] % = \nonumber \\
  =  \frac{1}{6} \int^{ \beta }_{0} d\tau  \left[  \Deltad ^3
  (\tau , 0) -  \Deltad ^3 (\tau , \beta ) \right] = \frac{\beta}{12} .
\label{wbn@9*}\end{eqnarray}
On the other hand, we apply
Eq.~(\ref{wbn@p7})
and perform two regular integrals,  reducing
$I_{14}$ to a form containing an
undefined integral
over a product of distributions:
\begin{eqnarray}
 I_{14} & = &  \int_0^\beta  \int_0^\beta  d\tau \, d\tau '\, \dDelta (\tau
,\tau ')
    \Deltad (\tau ,\tau ')  \delta (\tau -\tau ') %\nonumber \\&&
    - \frac{1}{ \beta } \int_0^\beta  \int_0^\beta  d\tau \, d\tau '
   \,  \dDelta  (\tau ,\tau ')  \Deltad (\tau ,\tau ')
    \nonumber \\
 & = &  \int_0^\beta  \int_0^\beta  d\tau \, d\tau ' \left[
-\frac{1}{4}\epsilon^2
   (\tau -\tau ')  \delta (\tau -\tau ') \right]  +  \int^{ \beta }_{0}
  d\tau\, \dDelta ^2 (\tau ,\tau ) + \frac{ \beta }{12} \nonumber \\
 & = &  \beta  \left[- \frac{1}{4} \int  d \tau \,  \epsilon ^2 (\tau )
     \delta (\tau ) + \frac{1}{6}\right].
\label{wbn@10*}\end{eqnarray}
A third, mixed  way of evaluating
$I_{14}$
employs one  partial integration  as in
the first line of Eq.~(\ref{wbn@9*}), then
the equation of motion (\ref{wbn@eom})
to  reduce $I_{14}$ to yet another form
\begin{eqnarray}
 I_{14} & =  & \frac{1}{2}  \int_0^\beta  \int_0^\beta  d\tau \, d\tau '
\Deltad ^2
    (\tau ,\tau ')  \delta (\tau -\tau ') = \nonumber \\
& = &  \frac{1}{8} \int_0^\beta  \int_0^\beta  d\tau \, d\tau '  \epsilon ^2
(\tau -\tau ')
   \delta  (\tau -\tau ') + \frac{1}{2} \int^{ \beta }_{0}
  d\tau  \dDelta ^2 (\tau ,\tau ) \nonumber \\
& = &  \beta  \left[\frac{1}{8} \int d\tau\,  \epsilon^2 (\tau )
    \delta (\tau ) + \frac{1}{24}\right].
\label{wbn@11*}\end{eqnarray}
We now see that if we
set
\begin{equation}
\int d\tau  [\epsilon (\tau )]^2 \, \delta (\tau )\equiv
 \frac{1}{3}
\label{wbn@6*}\end{equation}
the last two results (\ref{wbn@11*}) and
(\ref{wbn@10*}) coincide with the first in Eq.~(\ref{wbn@9*}).
The definition
(\ref{wbn@6*}) is obviously consistent with
partial integration
if we insert $ \delta(\tau )=\dot  \epsilon(\tau )/2$:
\begin{equation}
\int d\tau  [\epsilon (\tau )]^2 \, \delta (\tau )=
 \frac{1}{2}\int d\tau  [\epsilon (\tau )]^2 \, \dot \epsilon  (\tau )
= \frac{1}{6}\int d\tau  \frac{d}{d\tau }  [\epsilon (\tau )]^3
= \frac{1}{3}.
\label{wbn@n6*}\end{equation}
While the integration rule (\ref{wbn@6*})
is
consistent with partial integration and equation of
motion, it is incompatible
with the requirement
of coordinate independence. This can be seen from
the discrepancy
between the resulting value $I_{14} = \beta/12$
and the necessary
(\ref{wbn@I12}). This
 discrepancy was compensated in Refs.~\cite{5,6,7,8,9,10,11,12,13}
by adding the above-mentioned
noncovariant term to the classical
action.

A similar problem appears
with the other Feynman integral
(\ref{wbn@8*}). Applying first
Eq.~(\ref{wbn@p7}) we obtain
\begin{eqnarray}
      I_{15} = \int_0^\beta  \int_0^\beta  d\tau \, d\tau '  \Delta (\tau ,\tau
')
 \delta ^2 (\tau -\tau ') - \frac{2}{ \beta } \int^{ \beta }_{0}
  d\tau   \Delta (\tau ,\tau ) + \frac{1}{ \beta ^2} \int_0^\beta  \int_0^\beta
    d\tau  d\tau '  \Delta (\tau ,\tau ').
\label{wbn@12*}\end{eqnarray}
For the integral
containing the square of the $ \delta $-function
 we must postulate the integration rule
\begin{equation}
  \int d\tau  \left[  \delta (\tau )\right] ^2 f(\tau ) \equiv
  \delta (0) f(0)
\label{wbn@4*}\end{equation}
to obtain a divergent term
\begin{equation}
  I_{15}^{\div} =  \delta (0) \int^{ \beta }_{0} d\tau   \Delta
  (\tau ,\tau ) =  \delta  (0) \frac{ \beta ^2}{6}.
\label{wbn@13*}\end{equation}
proportional to $ \delta(0)$
compensating a similar term
from the measure.
The remaining integrals in (\ref{wbn@12*}) are
finite
and yield the regular part of $I_{15}$
\begin{equation}
 I_{15}^{R} = - \frac{ \beta }{4}.
\label{wbn@14*}\end{equation}
In another  calculation of  $I_{15}$,
we
first  add and subtract the UV divergent term,
writing
\begin{equation}
 I_{15} = \int_0^\beta  \int_0^\beta  d\tau\,  d\tau '  \Delta (\tau ,\tau ')
\left[
 \dDeltad  ^2 (\tau ,\tau ') -  \delta ^2 (\tau -\tau ')
  \right]  +  \delta (0) \frac{ \beta ^2}{6}.
\label{wbn@15*}\end{equation}
Replacing $ \delta ^2 (\tau -\tau ')
$ by the square of
the left-hand side of
the equation of motion
(\ref{wbn@eom}), and integrating the terms in  brackets
by parts, we obtain
\begin{eqnarray}
 I_{15}^{R} & =& \int_0^\beta  \int_0^\beta  d\tau\,  d\tau '  \Delta  (\tau
,\tau ')
  \left[ \dDeltad ^2 (\tau ,\tau ') -
   \Deltadd\, ^2 (\tau ,\tau ') \right]  \nonumber \\
& =& \int_0^\beta  \int_0^\beta  d\tau \, d\tau '  \left [ -  \dDelta(\tau
,\tau ')
    \Deltad (\tau ,\tau ')
 \dDeltad (\tau ,\tau ')
 - \Delta(\tau ,\tau ')
\Deltad (\tau ,\tau ')
\,\ddDeltad (\tau ,\tau ') \right ]\nonumber \\
     &-& \int_0^\beta  \int_0^\beta  d\tau \, d\tau '  \left [-\Deltad^2 (\tau
,\tau ')
\Deltadd (\tau ,\tau ') -
  \Delta (\tau ,\tau ')\Deltad (\tau ,\tau ')\Deltaddd  (\tau ,\tau ')\right ]
 \nonumber\\
 &=&
- I_{14} + \int_0^\beta  \int_0^\beta  d\tau \, d\tau '  \Deltad ^2  (\tau
,\tau ')
  \Deltadd (\tau ,\tau ')
=- I_{14} - \beta /6.
\label{wbn@16*}\end{eqnarray}
The value of the last integral
follows from
 partial integration.

For  a third evaluation of $I_{15}$
we  insert the equation of motion (\ref{wbn@eom})
and bring the last integral in the fourth line of (\ref{wbn@16*}) to
\begin{equation}
- \int_0^\beta  \int_0^\beta  d\tau \,d\tau '  \Deltad ^2 (\tau ,\tau ')
 \delta (\tau -\tau ')
  = - \beta  \left[\frac{1}{4} \int d\tau\,
\epsilon^2 (\tau )  \delta (\tau )
 + \frac{1}{12}\right].
\label{wbn@17*}\end{equation}
All three ways of calculation lead to the same result
$I_{15}^{R} = - \beta/4$
using the rule
 (\ref{wbn@6*}).
 This, however, is  again
in disagreement with the
coordinate-invariant value in Eq.~(\ref{wbn@I12}).
Note that both integrals $I_{14}$ and $I_{15}^{R}$ are
too large by a factor $2$ with respect to the necessary
(\ref{wbn@I12}) for coordinate invariance.

How can we save
coordinate invariance
while maintaining the
equation of motion and partial integration?
The direction in which the answer lies  is
suggested
by the last line of Eq.~(\ref{wbn@11*}):
we must find a consistent way
to have
 an integral
\begin{equation}
 \int d\tau  \left[ \epsilon (\tau ) \right] ^2  \delta (\tau ) = 0,
\label{wbn@5a*}\end{equation}
instead of (\ref{wbn@6*}),
which means that we need a
reason
for forbidding
the application of
partial integration to this singular integral.
For the calculation at the infinite time interval,
this problem was solved
in our previous papers \cite{2,3,3a}
with the help of the dimensional regularization,
carried to higher
orders
in Refs.~\cite{14,15}).
The extension of our rules
to the  short-time
amplitude considered here is straightforward.
It can be done
without performing
any of the  cumbersome
calculations in $1-\varepsilon$-dimension.
We must only keep track of the essential features
of the structure of the
Feynman integrals in arbitrary dimensions.
For this we
continue the imaginary time
coordinate $\tau $ to
a $d$-dimensional spacetime vector
$\tau \rightarrow x^\mu = (\tau , x^1, \dots, x^{d-1})$.
In $d=1-\varepsilon-$ dimensions, the correlation function reads
\begin{equation}
 \Delta (\tau, {\bf x};\,\tau',{\bf x}') = \int
 \frac{d^\varepsilon k}{(2\pi)^\varepsilon}
 e^{i{\sbf k}({\sbf x} - {\sbf x}')}
 \Delta_{\omega} (\tau,\tau').
\label{wbn@n1}\end{equation}
Here the extra $\varepsilon$-dimensional space coordinates ${\bf x}$
are assumed to live on
 infinite axes with
translational
invariance along all directions.
Only
 the
 $\tau$-coordinate
lies in a finite interval $0\leq\tau\leq\beta$,
with Dirichlet boundary conditions
for (\ref{wbn@n1}). The one-dimensional correlation function
$
 \Delta_{\omega} (\tau,\tau')
$
in the integrand has a mass
$\omega = |{\bf k}|$.
It is the Green function
on the finite $\tau$-interval
\begin{equation}
 - \ddDelta_{\omega} (\tau,\tau')+\omega^2
 \Delta_{\omega} (\tau,\tau') = \delta(\tau - \tau'),
\label{wbn@n2}\end{equation}
satisfying the Dirichlet boundary conditions
\begin{equation}\label{wbn@n3}
  \Delta_{\omega} (0,\tau) =  \Delta_{\omega} (\beta,\tau)
   = 0 .
\end{equation}
Explicitly, it reads \cite{PI}
\begin{equation}\label{wbn@n4}
 \Delta_{\omega} (\tau,\tau')= \frac{\sinh \omega
 (\beta - \tau_>)\,\sinh \omega\tau_<}{\omega \sinh\omega\beta}\,,
\end{equation}
where $\tau_>$ and $\tau_<$ denote the larger and smaller of the
imaginary times $\tau$ and $\tau'$, respectively.

In $d$ dimensions, the equation of motion (\ref{wbn@eom}) becomes a scalar
field equation of the Klein-Gordon type.
Using Eq.~(\ref{wbn@n2}), we obtain
\begin{eqnarray}\label{wbn@n5}
{}_{\mu\mu}\Delta (\tau,{\bf x};\,\tau',{\bf x}')
&=&\Delta_{\mu\mu} (\tau,{\bf x};\,\tau',{\bf x}')=
 \ddDelta (\tau,{\bf x};\, \tau',{\bf x}') + \,\,
{}_{{\bf x}{\bf x}}\Delta (\tau,{\bf x}; \,\tau',{\bf x}')
\nonumber\\
 &=& \int \frac{d^\varepsilon  k}{(2\pi)^\varepsilon}
e^{i{\sbf k}({\sbf x} - {\sbf x}')}
 \left[\ddDelta_{\omega}(\tau,\tau')-\omega^2\Delta_{\omega}
 (\tau,\tau')\right]=\nonumber\\
 &=& -\,\delta (\tau-\tau')\, \delta^{(\varepsilon)} ({\bf x} - {\bf x}')
 = -\,\delta^{(d)} (x-x').
\end{eqnarray}

The important observation is now that
for $d$ spacetime dimensions,
perturbation expansion of the
 path integral yields for the second correlation function
$\dDeltad (\tau,\tau')$ in Eqs. (\ref{wbn@7*}) and (\ref{wbn@8*})
the extension  ${}_\mu \Delta_\nu (x,x')$.
This function differs from
the contracted function
${}_\mu\Delta_\mu (x,x')$, and
from
${}_{\mu\mu}\Delta (x,x')$
which satisfies
the field equation
(\ref{wbn@n5}). In fact, all correlation functions
 $\dDeltad (\tau,\tau')$ encountered in the
diagrammatic expansion which have different time arguments always
have the $d$-dimensional extension
 ${}_\mu\Delta_\nu (x,x')$. An important exception
is the
correlation functions at {\em equal\/} times
 $\dDeltad (\tau,\tau)$ whose $d$-dimensional
extension is always ${}_\mu \Delta_\mu (x,x)$, which
satisfies the equation (\ref{wbn@p7}) in the $\varepsilon\to 0$-limit.
Indeed, it follows from Eq.~(\ref{wbn@n1}) that
\begin{equation}\label{wbn@n6}
  {}_\mu \Delta _\mu (x,x) = \int \frac{d^\varepsilon
  k}{(2\pi)^\varepsilon}
  \left[\dDeltad_{\omega} (\tau,\tau) + \omega^2 \,
  \Delta_{\omega} (\tau,\tau)\right].
\end{equation}
With the help of Eq.~(\ref{wbn@n4}), the integrand
in Eq.~(\ref{wbn@n6}) can be brought to
\begin{equation}\label{wbn@n7}
  \dDeltad_{\omega} (\tau,\tau) + \omega^2
  \Delta_{\omega} (\tau,\tau) = \delta(0) - \frac{\omega \cosh\omega
(2\tau-\beta)}
  {\sinh\omega\beta}\,.
\end{equation}
Substituting this into Eq.~(\ref{wbn@n6}), we obtain
\begin{equation}\label{wbn@n8}
  {}_\mu \Delta_\mu (x,x) = \delta^{(d)} (x,x) - I^\varepsilon\,.
\end{equation}
The integral $I^\varepsilon$ is
 calculated as
 follows
\begin{eqnarray}\label{wbn@n9}
I^\varepsilon &= &\int \frac{d^\varepsilon
k}{(2\pi)^\varepsilon} \frac{\omega \cosh\omega
(2\tau-\beta)}{\sinh\omega\beta} =
%\nonumber\\ &=&
 {1\over\beta}
\frac{S_\varepsilon}{(2\pi\beta)^\varepsilon} \int^\infty_0
dzz^\varepsilon \frac{\cosh z(1-{2\tau/\beta})}{\sinh z} \\ &=&
{1\over \beta} \frac{S_\varepsilon}{(2\pi\beta)^\varepsilon}
\frac{\Gamma (\varepsilon+1)}{2^{\varepsilon+1}}
\big[\zeta(\varepsilon+1,1-{\tau\over\beta}) +
\zeta(\varepsilon+1,{\tau\over\beta}) \big],\nonumber
\end{eqnarray}
where $S_\varepsilon=2\pi^{\varepsilon/2}/\Gamma (\varepsilon/2)$
is the surface of unit sphere in $\varepsilon$ dimension, and $\Gamma (z)$ and
$\zeta(z,q)$ are gamma and zeta functions, respectively.
For small $\varepsilon\to 0$,
they have the limits
$\zeta(\varepsilon+1,q)\to 1/\varepsilon- \psi(q)$,
and  $\Gamma (\varepsilon/2)\to 2/\varepsilon$, so that
$I^\varepsilon\to 1/\beta$,
proving that
the $d$-dimensional equation (\ref{wbn@n8})
at coinciding arguments reduces indeed to
the one-dimensional equation (\ref{wbn@p7}).
The explicit $d$-dimensional form will
never be needed, since we can always treat
$ {}_\mu  \Delta _\mu(x,x)$
as purely one-dimensional  objects
$\dDeltad(\tau , \tau )$, which can in turn
be replaced everywhere by the right-hand side
$ \delta (0)-1/ \beta $
 of  (\ref{wbn@p7}).

We now show that
by carefully keeping track
of the different contractions of the derivatives,
we obtain
a consistent
calculation scheme
which yields results
equivalent to assuming   integration rule
(\ref{wbn@5a*})
in the calculation of
$I_{14}$ and $I_{15}$,
thus ensuring
coordinate independence.
The integral (\ref{wbn@7*})
for $I_{14}$ is extended to
\begin{equation}
  I_{14}^{d} = \int \int d^dx \, d^dx' {}_\mu \Delta  (x,x')  \Delta _ \nu
   (x,x')\,_\mu  \Delta _ \nu  (x,x'),
\label{wbn@18*}\end{equation}
and the different derivatives on $\,_\mu  \Delta _ \nu  (x,x')$
prevent us from applying
the field equation (\ref{wbn@n5}), in contrast to the
one-dimensional calculation.
We can, however, apply partial integration
as in the first line of Eq.~(\ref{wbn@9*}),
and arrive at
\begin{eqnarray}
  I_{14}^{d} & = &  - \frac{1}{2} \int \int d^dx \, d^dx'  \Delta _ \nu ^2
(x,x')
     \Delta _{\mu\mu} (x,x').% \mathop{\rightarrow}^{d\rightarrow 1}
 %-\frac{1}{2} \int \int d\tau \, d\tau '  \Deltad ^2 (\tau ,\tau ')
 % \Deltadd (\tau, \tau ').
\label{wbn@19*}\end{eqnarray}
In contrast to the one-dimensional expression
(\ref{wbn@9*}),
a further partial integration is impossible. Instead, we
apply the field equation  (\ref{wbn@n5}), go back to one dimension,
and apply the
integration rule (\ref{wbn@5a*})
as in Eq.~(\ref{wbn@11*}) to obtain the correct result
$I_{14} =  \beta /24$ guaranteeing coordinate invariance.

  The Feynman integral (\ref{wbn@8*}) for $I_{15}$ is treated likewise.
Its $d$-dimensional extension is
\begin{equation}
 I_{15}^{d} =
 \int \int d^dx \, d^dx'  \Delta (x,x')
\left[ {}_\mu \Delta _ \nu (x,x') \right] ^2 .
\label{wbn@20*}\end{equation}
The different derivatives
 on ${}_\mu \Delta _ \nu (x,x')$  make
it impossible to apply
a dimensionally extended version of equation
(\ref{wbn@p7})
as in Eq.~(\ref{wbn@12*}).
We can, however, extract the UV
divergence
 as in Eq.~(\ref{wbn@15*}),
 and perform a partial integration on the finite part
which brings it to  a dimensionally  extended version
of Eq.~(\ref{wbn@16*}):
\begin{equation}
 I_{15}^{R}  = - I_{14} + \int d^dx \, d^dx'  \Delta _ \nu ^2 (x,x')
    \Delta _{\mu\mu}   (x,x').
\label{wbn@21*}\end{equation}
On the right-hand side
we use
the field equation
(\ref{wbn@n5}), as in  Eq.~(\ref{wbn@17*}),
return to $d=1$, and
use the rule (\ref{wbn@5a*})
to obtain
the result $I_{15}^{R} = - I_{14} -  \beta /12 = - \beta /8$, again
guaranteeing coordinate independence.

Thus, by keeping only track
of a few essential properties
of the theory
in $d$ dimensions
we indeed obtain a simple
consistent
procedure for calculating singular
Feynman integrals. All
results obtained in  this way
ensure
 coordinate independence. They
agree with what we would obtain
using the
one-dimensional integration rule
  (\ref{wbn@5a*}) for the product of
two $ \epsilon $- and one $ \delta $-distribution.

Our procedure
gives us unique rules telling us where we are allowed
to apply partial integration
and the equation of motion
in one-dimensional expressions.
Ultimately,
all integrals
are brought
to
a
regular form, which can be
continued back
to
one time dimension for a
direct evaluation.
This procedure
is obviously  much  simpler than
the previous explicit calculations in $d$ dimensions
with the limit $d \rightarrow 1$ taken at the end.

We now apply this procedure
to the perturbation expansion
of the short-time amplitude
of a  free particle
in curvilinear coordinates.

\section{Perturbation Expansion}

A free point particle of unit mass
has the action
\begin{equation}
{\cal A}_ {0} [\qn]   =
\frac{1}{2}\,\int_0^ \beta  d \tau \,\dot \qn ^2(\tau ).
\label{wbn@m1}\end{equation}
The amplitude
$ (0\, \beta |0\,0)_0 $
is given by the Gaussian path integral
\begin{equation}
  (0\, \beta |0\,0)_0  = \int  {\cal D} \qn  (\tau)\,
e^{-{\cal A}_{0} [\qn ]}
= e^{- (1/2){\rm T\!r} \log (-\partial^2 )} =\left[ 2\pi \beta \right] ^{-1/2}
.\label{wbn@mq2}\end{equation}
A coordinate transformation $\qn(\tau ) = f(q(\tau ))$
brings the action (\ref{wbn@m1}) to the form
\begin{eqnarray}
 {\cal A} [q] = \frac{1}{2}
 \int^{ \beta }_{0} d\tau    \, g (q (\tau )) \dot q^2 (\tau ),
\label{wbn@1**}\end{eqnarray}
 where $g(q) = f'{}^2 (q)$. The measure ${\cal D} \qn (\tau )
 \equiv \prod_{\tau } d\qn (\tau )$ transforms as follows:
\begin{equation}
   {\cal D} \qn(\tau )\equiv \prod_{\tau } d\qn (\tau ) = J \,
  \prod_{\tau } dq (\tau ) \equiv  J \, {\cal D} q(\tau ),
\label{wbn@2**}\end{equation}
where $J$ is the Jacobian of the coordinate transformation
\begin{equation}
  J = e^{(1/2 ) \delta (0) \int_{0}^{ \beta } d\tau  \log
  g ( q (\tau ))}.
\label{wbn@3**}\end{equation}
 Thus the transformed path integral (\ref{wbn@mq2})
takes precisely the form (\ref{wbn@@1a}),
with the total action in the exponent
\begin{equation}
 {\cal A}_{\rm tot} [q] = \int^{ \beta }_{0} d\tau
   \left[ \frac{1}{2} g (q(\tau )) \dot q^2 (\tau ) - \frac{1}{2}
   \delta (0) \log g (q(\tau ))\right] .
\label{wbn@4**}\end{equation}
This is decomposed into a free part
\begin{equation}
 {\cal A}_{0} [q] = \frac{1}{2} \int_{0}^{ \beta } d\tau \, \dot q^2 (\tau )
\label{wbn@5**}\end{equation}
and an interacting part
\begin{equation}
 {\cal A}_{\inter} [q] = \int_{0}^{ \beta } d\tau  \frac{1}{2}
  [g(q)-1] \dot q^2 - \int_{0}^{ \beta } d\tau  \frac{1}{2}
  \delta (0) \left\{ [g(q) - 1] - \frac{1}{2} [g(q)-1]^2 + \dots
 \right\}      .
\label{wbn@6**}\end{equation}
The path integral (\ref{wbn@mq2})
is now formally
defined by the
perturbation expansion
\begin{eqnarray}
 \langle 0,  \beta  \vert 0,0\rangle & = &  \int {\cal D} q (\tau )
     e^{{\cal A}_{0}  [q] - {\cal A}_{\inter} [q] }
% \nonumber \\ & = &
=  \int {\cal D}  q (\tau ) e ^{-{\cal A}_{0} [q]} \left( 1 -
    {\cal A}_{\inter} + \frac{1}{2} {\cal  A}^2_{\inter} -
\dots \right)
 \nonumber \\
 & = &  (2 \pi  \beta )^{-1/2} \left[ 1 - \langle {\cal A}_{\inter}
          \rangle + \frac{1}{2} \langle {\cal A}^2_{\inter} \rangle -
    \dots \right],
\nonumber \\
 & = &  (2 \pi  \beta )^{-1/2} e^{ - \langle {\cal A}_{\inter}
          \rangle_c + \frac{1}{2} \langle {\cal A}^2_{\inter} \rangle_c -
    \dots } ,
\label{wbn@7**}\end{eqnarray}
with the harmonic expectation values
\begin{equation}
  \langle \dots \rangle = (2 \pi  \beta )^{1/2} \int {\cal D}
      q (\tau ) ( \dots ) e^{- {\cal A}_0 [q]},
\label{wbn@8**}\end{equation}
and their cumulants
$ \langle {\cal A}_{\inter}^2 \rangle_c = \langle {\cal A}^2 _{\inter}
\rangle-  \langle {\cal A}_{\inter} \rangle ^2, \dots~$
containing only connected diagrams.
If our calculation procedure
respects coordinate independence,
all expansion terms must vanish
to yield the trivial exact results (\ref{wbn@mq2}).
As an example we shall consider
the coordinate transformation
\begin{eqnarray}
x= f(q) = q - \frac{1}{3} \varepsilon q^3 + \frac{1}{5} \varepsilon^2
 q^5 - \dots ,
\label{wbn@9**}\end{eqnarray}
 such that
\begin{equation}
 g(q) = f'^2(q)=1 - 2 \varepsilon q^2
+ 3 \varepsilon ^2 q ^4 -4\varepsilon ^3
	    q ^6 + \dots ,
\label{wbn@10**}\end{equation}
 where $\varepsilon$ is a smallness parameter. Substituting
 Eq.~(\ref{wbn@10**}) into Eq.~(\ref{wbn@6**})
yields
up to second order in $\varepsilon$:
\begin{eqnarray}
  {\cal A}_{\inter} [q] & = & \int^{ \beta }_{0} d\tau
    \left\{ \left[ - \varepsilon q ^2 (\tau ) + \frac{3}{2} \varepsilon^2
   q ^4 (\tau ) \right]  \dot q^2 (\tau )
%\right. \nonumber \\  & &\left.
  -  \delta (0) \left[ - \varepsilon q^2 (\tau ) +
   \frac{1}{2} \varepsilon^2 q ^4 (\tau ) \right] \right\} .
\label{wbn@11**}\end{eqnarray}

We shall now calculate order by order in $ \varepsilon $
the expansion terms contributing to the square bracket in
the second line of
Eq.~(\ref{wbn@7**}).

\section{Diagrams}
To first order in $\varepsilon $, the
 square bracket in
the second line of
Eq.~(\ref{wbn@7**}) receives
a contribution from the expectation values
of the linear terms in $ \varepsilon $  of the interaction (\ref{wbn@11**}):
\begin{equation}
- \langle {\cal A}_{\inter}^{{\rm lin~in}\,\varepsilon }
          \rangle =
 \int^{ \beta }_{0} d\tau\,
\big\langle
 \varepsilon q ^2 (\tau )
  \dot q^2 (\tau )
- \delta (0)  \varepsilon q^2 (\tau )
\big\rangle.
\label{wbn@@}\end{equation}
Thus
there
exists only three diagrams,
two originating from the kinetic term and
one from the Jacobian action:
\begin{equation}
 \,\varepsilon\hspace{0mm}\raisebox{-1mm}{\mbox{\input 6.tps }} + 2\,\varepsilon\,
       \hspace{-0mm}\raisebox{-.90mm}{\mbox{\input infh.tps }}
 - \varepsilon\,\delta (0) \hspace{0mm}\raisebox{-1mm}{\mbox{\input 0dot.tps }} .
\label{wbn@f1}\end{equation}

To order $\varepsilon^2$,
we need to calculate only connected diagrams
contained in the term
$\langle {\cal A}_{\inter}^2 \rangle/2$ in (\ref{wbn@7**}), all
disconnected ones being
obtainable from the
cumulant relation
$ \langle {\cal A}_{\inter}^2 \rangle = \langle {\cal A}^2 _{\inter}
\rangle_c  +  \langle {\cal A}_{\inter} \rangle ^2.$
We distinguish several contributions.

First, there are two local three-loop diagrams
and one two-loop local diagram
coming
from
the kinetic term and the Jacobian of the interaction
(\ref{wbn@11**}), respectively:
\begin{eqnarray}
 &&  \left(-\frac{3}{2}\varepsilon ^2\right)  \,\Bigg[\,\,\,
 3 \hspace{0mm}\raisebox{-3.2mm}{\mbox{\input 7.tps }} +
\,12
 \hspace{-0mm}\raisebox{-2.4mm}{\mbox{\input cloverh.tps }}
- \, \delta (0)\,\,
\hspace{-27mm}\raisebox{-11.57mm}{\mbox{\input inf.tps }}
{}~\,~~
{}~~~~~~~~~
{}~~~~~~~~~
\,\,\, \Bigg]\,.
\label{wbn@f2}\end{eqnarray}
{}~\\[-1.1cm]
We call a diagram local if it involves
only equal-time Wick contractions.

 The Jacobian part of the
action (\ref{wbn@11**}) contributes further
the nonlocal diagrams:
\begin{eqnarray}
 &&\!\!\!\!\!\!
 \frac{\varepsilon^2}{2!}\bigg\{
 2\,\delta^2 (0) \!\!\hspace{0mm}\raisebox{-1mm}{\mbox{\input 0dotdot.tps }}
 \!\!- 4\,\delta (0) \big[\!\!
\hspace{0mm}\raisebox{-1mm}{\mbox{\input 6dot.tps }}\!\!+\!\!\!\!\!\!
\hspace{0mm}\raisebox{-1mm}{\mbox{\input 6pdot.tps }}
\!\!+ 4\,\!\!\!\hspace{0mm}\raisebox{-2mm}{\mbox{\input infdoth.tps }}\big]\bigg\}.
\label{wbn@f3}\end{eqnarray}
The remaining diagrams
come from
the kinetic term of
the interaction  (\ref{wbn@11**}) only.
They  are either of the
three-bubble type, or of the watermelon
type, each with all possible combinations
of the three line types (\ref{wbn@prop}):
The sum of all three-bubbles diagrams
is
\begin{eqnarray}
 \!\,\frac{\varepsilon^2}{2!}\big[ 4\!\!\!\hspace{0mm}\raisebox{-1.2mm}{\mbox{\input 8.tps }}
\!\!\!\hspace{1pt}+\!\!\!\,\,\,2~\,\!\!\!\!\!\hspace{0mm}\raisebox{-1.2mm}{\mbox{\input 9.tps }}
{}~\!\!\!+\!\!\!\,\,\,2\!\!\!\!\!\!\,\,\,\hspace{0mm}\raisebox{-1.2mm}{\mbox{\input 10.tps }}
%\nonumber \\[-0mm]&&~~~~~
\!+\!\,16\,\!
 \!\!\!\!\hspace{-0mm}\raisebox{-3mm}{\mbox{\input thr1h.tps }}
\! +\!16\!\!\!\!
 \hspace{-0mm}\raisebox{-3mm}{\mbox{\input thr1hh.tps }}
+\!16\, \!\!\!\!\!
 \hspace{-0mm}\raisebox{-3mm}{\mbox{\input thr1hhh.tps }} \!+16\,\!\!\!\!\!
 \hspace{-.0mm}\raisebox{-3mm}{\mbox{\input thr1hhhh.tps }}
 \big]     .
\label{wbn@f4}\end{eqnarray}

The watermelon-type diagrams
contribute
\begin{eqnarray}
& &\!\!\!\!\!\!\!\!\!\!
{\!\!\frac{\varepsilon^2}{2!}\, 4 \,\bigg[\!
\!\!\!\hspace{0mm}\raisebox{-2mm}{\mbox{\input 11.tps }}
\!+\, 4\!\!\!\hspace{0mm}\raisebox{-1.95mm}{\mbox{\input 12.tps }}
\!+\!\!\!\! \hspace{0mm}\raisebox{-1.9mm}{\mbox{\input 13.tps }}
\bigg].   \!\!\!\!\!        }
\label{wbn@f5}
\end{eqnarray}
{}~\\[-.4cm]

\section{Path Integral in Curved Space}

Before we start evaluating
the above Feynman diagrams,
we observe
that the same diagrams
appear
if we
define path integral
in
a higher-dimensional
target space $q^{i}$.
The generalization of the formal expression
(\ref{wbn@4**})
is obvious: we replace $g(q)$ by $g_{ij} (q)$
in the kinetic term, and by $g(q) \rightarrow \det
(g_{ij}(q))$ in the measure, where $g_{ij} (q)$
is the metric induced by
the
coordinate transformation
from cartesian
to curvilinear coordinates.
In a
further step, we shall
also consider
$g_{ij} (q)$
more generally as a metric in a
curved space,
which can be reached from
a flat space only by a nonholonomic coordinate transformation
\cite{NH}.
It was shown in the textbook
\cite{PI}
that under nonholonomic coordinate transformations,
the measure
of a time-sliced  path integral transforms
from the flat-space form
$\prod _n{d}x_n$ to
$\prod_n{d}q \sqrt{g_n}\exp( \Delta t R_n/6) $,
which has the consequence that
the amplitude satisfies a Schr\"odinger equation
with
the pure Laplace-Beltrami operator
in the kinetic Hamiltonian,
containing  no extra $R$-term.
Here we shall see that
a similar thing must happen
for perturbatively defined path integrals,
where the nonholonomic transformation
must carry the flat-space
measure
\begin{equation}
{\cal D}x\rightarrow
{\cal D}q \,\sqrt{g}\,\exp\left(\int_{0}^{ \beta } d\tau\, R/8\right) .
\label{wbn@@R8}\end{equation}
The proof of this
rather technical issue is relegated to a separate paper.

For $n$-dimensional manifolds
with a general metric
$g_{ij} (q)$
we make use of the coordinate
invariance to be proved
by the vanishing of the
expansion  (\ref{wbn@f1})--(\ref{wbn@f5}). This will allow us to
bring the metric to
the most convenient
 Riemann normal  coordinates.
Assuming $n$-dimensional manifold
to be a homogeneous space, as
in a standard nonlinear $\sigma$-model,
we expand
the metric and its determinant
in the normal coordinates
as follows
\begin{eqnarray}
  g_{ij}  (q) & = &   \delta _{ij} + \varepsilon\,\frac{1}{3}\,
 R_{ik_1 jk_2}\, q^{k_1} q^{k_2}
 + \varepsilon^2 \,
  \frac{2}{45}  R_{k_1 j k_2}{} ^l \,
 R_{k_3 i k_4 l}\, q^{k_1} q^{k_2} q^{k_3}  q^{k_4}
 + \dots ,\label{wbn@12**a} \\
 g(q) & = & 1 - \varepsilon\,\frac{1}{3} \, R_{ij}\, q^i q ^j
 + \varepsilon^2 \,
  \frac{1}{18}\left( R_{ij}\,R_{kl} + \frac{1}{5}\,
 R_{i n j }{} ^m \, R_{k m l}{}^n
\right) q^{i} q^{j} q^{k}  q^{l}
 + \dots~.
\label{wbn@12**}\end{eqnarray}
In our conventions, the Riemann and Ricci tensors are
$R_{ijk}{}^l = \partial_i  \Gamma _{jk}{}^l - \dots , R_{jk} =
R_{ijk}{}^i$,
and the curvature $R = R_i{}^i$ has the positive sign
for a sphere.
 The expansions
(\ref{wbn@12**a}) and (\ref{wbn@12**})
have obviously a similar power structure
in $q^i$  as the previous expansion
(\ref{wbn@10**}).

In  normal coordinates,
the interaction
(\ref{wbn@6**}) becomes,
up to order $\varepsilon^2$:
\begin{eqnarray}
\!\!\!\!\!\!\!\!\!\!\!\!\!\!\!{\cal A}_{\inter} [q] & = &
\int^{ \beta }_{0} d\tau  \Big\{\Big[
 \varepsilon \frac{1}{6}  R_{ik jl}\, q^{k} q^{l}
 + \varepsilon^2
  \frac{1}{45}  R_{m j n}{} ^l \,
 R_{r i s l}\, q^{m} q^{n} q^{r}  q^{s} \Big]
  \dot q^i \dot q ^j   \nonumber \\
 & &~ ~~~~~~~~\,+ \varepsilon \frac{1}{6} \delta (0) R_{ij}\, q^i q ^j +
   \varepsilon^2 \frac{1}{180}  \delta (0) R_{imj}{} ^n \,
	   R_{knl}{} ^{m} \,  q^i q^j q^k q^l \Big\},
\label{wbn@13**}\end{eqnarray}
with the same powers of $q^i$ as in Eq.~(\ref{wbn@11**}).
The interactions  (\ref{wbn@11**}) and (\ref{wbn@13**})
yield the same diagrams
in the perturbation expansions in powers
of $\varepsilon$.
In one dimension and with the trivial vertices
in the interaction (\ref{wbn@11**}),
the sum
of all diagrams will be shown to
vanish
in the case of a flat space.
In a curved space with the more complicated
vertices proportional
to $R_{ijkl}$ and $R_{ij}$, the same
Feynman integrals
will yield a
nontrivial short-time
amplitude.
The explicit $R_{ijkl}$-dependence
coming from the interaction vertices
is easily identified in the diagrams: all bubbles
in (\ref{wbn@f3})--(\ref{wbn@f4}) yield results proportional to
$R_{ij}^2$, while the watermelon-like
diagrams (\ref{wbn@f5})
carry a factor   $R_{ijkl}^2$.
In our
previous work \cite{2,3,3a},
all
integrals were calculated
in
$d$ dimensions,
taking the limit
$ d \rightarrow 1$ at the end.
In this way we confirmed
that the sum of all
Feynman diagrams
contributing to each order in $\varepsilon$ vanishes.
It is easy to verify that the same results
are
found using the
procedure
developed above.

\section{From Coordinate Independence to DeWitt-Seeley Expansion}

With the same procedure we now calculate
the first two terms in the
short-time expansion of the time-evolution
amplitude.  The results will be compared
with the similar expansion
obtained from
the operator
expression
for the
amplitude $e^{\beta D^2 \beta  /2}$
with
the Laplace-Beltrami operator $D^2 = g^{-1/2}\partial_i
g^{1/2}g^{ij}(q)\partial_j$
first derived by DeWitt\cite{DW}
(see also \cite{Seeley}):
\begin{eqnarray}\label{wbn@b1a}
(q,\beta\mid q',0) &\!\!=\!\!&
(q\mid e^{{\beta}D^2/2} \mid q') =
\frac{1}{ \sqrt{2\pi  \beta}^{n}}\,
 e^{- g_{ij}{\Delta q^i \Delta
q^j}/{2\beta}}\, \sum^\infty_{k=0} \beta^k a_k (q,q'),
\end{eqnarray}
with the expansion  coefficients
being for a homogeneous space
\begin{eqnarray}\label{wbn@sc}
a_0(q,q') &\!\!\equiv\!\! & \!1 + {1\over 12} R_{ij} \Delta q^i
\Delta q^j + \left(  {1\over 360} R^i{}_k {^j}{_l} R_{imjn}
+
%\nonumber\\&+&
 {1\over 288} R_{kl} R_{mn}\right)
 \Delta q^k
\Delta q^l  \Delta q^m \Delta q^n +
...\,,\nonumber\\
a_1 (q,q') &\!\!\equiv \!\!&\! {1\over 12} R \!+\!
\left({1\over 144} R~ R_{ij} \!+\!
{1\over 360} R^{kl} R_{kilj}\! + \!
% \nonumber\\&+&
{1\over 360} R^{kl
m}{_i} R_{klmj} - {1\over 180} R^n_i R_{nj}\right)
 \Delta q^i
\Delta q^j+ ...\,,\nonumber\\
a_2 (q,q') &\equiv & \!{1\over 288} R^2 +
{1\over 720} R^{ijkl} R_{ijkl} - {1\over 720} R^{ij} R_{ij}
 +  ...\,,
\end{eqnarray}
where $\Delta q^i \equiv (q-q')^i$.
For $ \Delta q^i = 0$ this simplifies to
\begin{eqnarray}\label{wbn@b1b2}
\!\!\!\!\!\!\!\!\!\!(0,\beta\mid 0,0) &=&
\frac{1}{ \sqrt{2\pi  \beta}^{n}}\,
 \bigg\{ 1+
{\beta\over 12} R(q)+\frac{ \beta ^2}{72}\left[
\frac{1}{4}R^2+\frac{1}{10}\left(
R^{ijkl}R_{ijkl}  -
R^{ij}R_{ij}
\right)\right] \bigg\}.
\label{wbn@@scp}
\end{eqnarray}
The derivation is sketched in
Appendix B.

For coordinate independence, the sum of the first-order
diagrams (\ref{wbn@f1}) has
to vanish.
Analytically, this amounts to the equation
\begin{equation}
 \int^{ \beta }_{0} d\tau  \left[\Delta (\tau ,\tau )
 \dDeltad (\tau ,\tau ) + 2 \dDelta ^2 (\tau ,\tau ) -
   \delta (0)  \Delta  (\tau ,\tau )\right] = 0.
\label{wbn@22}\end{equation}
In the $d$-dimensional extension,
the correlation function $\dDeltad (\tau ,\tau )$
at equal times
is the
limit
$d \rightarrow 1~$
of the contracted
correlation function ${}_\mu \Delta _\mu(x,x)$ which
satisfies the $d$-dimensional field equation~(\ref{wbn@n5}).
Thus
we can use Eq.~(\ref{wbn@p7}) to
replace
$\dDeltad (\tau ,\tau )$
by $ \delta (0) - 1/ \beta $.
This removes
the infinite factor $ \delta (0)$
in Eq.~(\ref{wbn@22})
coming from the measure.
The remainder is calculated directly:
\begin{equation}
\!\!\!\!\!\! \int_{0}^{ \beta } d\tau  \left[  - \frac{1 }{ \beta }\Delta (\tau ,\tau)
 + 2\, \dDelta ^2 (\tau ,\tau ) \right] =0.
% \frac{ \beta }{2} \epsilon^2 (0) = 0.
\label{wbn@1***}\end{equation}
This result is obtained without
subtleties, since by
Eqs.~(\ref{wbn@p4}) and (\ref{wbn@p5})
\begin{equation}
 \Delta (\tau ,\tau )= \tau - \frac{\tau ^2}{ \beta },~~~~
 \dDelta^2 (\tau ,\tau )=\frac{1}{4}- \frac{\Delta (\tau, \tau)}{ \beta }  ,
\label{wbn@@2a***}\end{equation}
whose integrals yield
\begin{eqnarray}
    \frac{1}{2 \beta }  \int^{ \beta }_{0} d\tau   \Delta (\tau ,\tau )=
 \int^{ \beta }_{0} d\tau \,\dDelta ^2 (\tau ,\tau ) =
   \frac{ \beta }{12}.
\label{wbn@2***}\end{eqnarray}
 The  same first-order diagrams (\ref{wbn@f1})
appear in curved space, albeit in different combinations:
\begin{eqnarray}
  && - \frac{1}{6} R \int^{ \beta }_{0} d\tau  \left[  \Delta  (\tau ,\tau )
 \dDeltad  (\tau ,\tau ) - \dDelta ^2 (\tau ,\tau )
 -  \delta (0)  \Delta (\tau ,\tau )\right],
\label{wbn@3***0}\end{eqnarray}
which is evaluated,
using the integrals (\ref{wbn@2***}),
to
\begin{eqnarray}
   \frac{1}{6} R \int^{ \beta }_{0} d\tau  \left[
 \frac{1 }{ \beta } \Delta (\tau ,\tau )
  + \dDelta ^2 (\tau ,\tau ) \right]  =    \frac{ \beta }{24} R.
\label{wbn@3***}\end{eqnarray}
%
%For a flat space in curvilinear coordinates the sum (\ref{wbn@3***})
%is again equal to zero in agreement with reparametrization
%invariance of the path integral.
This has to be supplemented
by a similar contribution
coming from the nonholonomically
transformed measure (\ref{wbn@@R8}).
Both terms together
yield
the  first-order DeWitt-Seeley expansion
\begin{equation}
(0\, \beta |0\,0)\equiv
\langle e^{\beta D^2 /2}\rangle =
\frac{1}{ \sqrt{2\pi  \beta}^{n}}\,
\left(1 +  \frac{ \beta }{12}R \right),
\label{wbn@fosc}\end{equation}
in agreement with (\ref{wbn@b1b2}).

We now turn to the evaluation of
the second-order diagrams.
The sum of the local diagrams (\ref{wbn@f2}) is given
by
\begin{equation}
 \sum (\ref{wbn@f2}) = - \frac{3}{2} \varepsilon^2 \int^{ \beta }_{0}
   d\tau  \left[ 3 \Delta ^2 (\tau ,\tau ) \dDeltad
 (\tau , \tau )
  + 12  \Delta (\tau ,\tau ) \dDelta ^2 (\tau ,\tau )
 -  \delta (0)  \Delta ^2 (\tau ,\tau ) \right] .
\label{wbn@29}\end{equation}
 Replacing $\dDeltad (\tau ,\tau )$ in Eq.~(\ref{wbn@29})
again by $  \delta (0) - 1/ \beta $,
and taking into account the
equality
\begin{equation}
\int_{0}^{ \beta } d\tau \,\Delta  (\tau ,\tau ) \left[
   \frac{1}{ \beta } \Delta (\tau ,\tau ) - 4 \dDelta ^2
 (\tau ,\tau )\right]  = 0
\label{wbn@4***}\end{equation}
following from Eq.~(\ref{wbn@@2a***}),
we find only the divergent term
\begin{equation}
 \sum(\ref{wbn@f2})  = \varepsilon^2 \left[ - 3  \delta (0) \int^{ \beta }_{0}
  d\tau   \Delta ^2 (\tau ,\tau ) \right]   = \varepsilon^2
 \left[  - \frac{ \beta ^3}{10}\,\delta (0)\right] .
\label{wbn@31}\end{equation}

  The sum of all bubbles diagrams (\ref{wbn@f3})--(\ref{wbn@f4})
resembles a Russian doll, where the partial
sums of different diagrams are embedded into each other.
Therefore, we begin the calculation with  the sum
(\ref{wbn@f3})
whose analytic form is
\begin{eqnarray}
 \!\!\!\! \sum (\ref{wbn@f3})
 & \!\!= \!\!& \frac{\varepsilon^2}{2}  \int_0^\beta  \int_0^\beta
     d\tau\,  d\tau ' \left\{ 2  \delta ^2 (0)  \Delta ^2 (\tau ,\tau ')
  \right. \nonumber \\
\!\!\!\!& & \!\!- 4 \left. \delta (0) \left[\Delta (\tau ,\tau ) \dDelta ^2
 (\tau ,\tau ') + 4 \dDelta  (\tau ,\tau )  \Delta (\tau ,\tau ') \dDelta
   (\tau ,\tau ') +  \Delta ^2 (\tau ,\tau ') \dDeltad
 (\tau , \tau )\right] \right\} .
\label{wbn@32}\end{eqnarray}
Inserting Eq.~(\ref{wbn@p7})
into the last equal-time term, we obtain
\begin{eqnarray}
  \sum (\ref{wbn@f3})& =& \frac{\varepsilon^2}{2} \int_0^\beta  \int_0^\beta
 d\tau\, d\tau '
\left\{-  2  \delta ^2 (0)  \Delta ^2 (\tau ,\tau ')
  \right. \nonumber \\
&&\!\!\!\!\left.-  4  \delta (0) \left[  \Delta (\tau ,\tau ) \dDelta ^2
 (\tau ,\tau ') +  4\,\dDelta (\tau ,\tau )  \Delta (\tau ,\tau ')
 \dDelta (\tau ,\tau ') -  \Delta ^2 (\tau ,\tau ')/ \beta\right]
 \right\} .
\label{wbn@5***}\end{eqnarray}
As we shall see below, the explicit evaluation of the
integrals in this sum is not necessary.
Just for completeness,
we
give
 the result:
\begin{eqnarray}
  \sum (\ref{wbn@f3}) & = & \frac{\varepsilon^2}{2}
   \left\{-  2  \delta ^2 (0) \frac{ \beta ^4}{90}
   - 4  \delta (0) \left[ \frac{ \beta ^3}{45} + 4 \frac{ \beta ^3}{180}
    - \frac{1}{ \beta } \cdot \frac{ \beta ^4}{90}\right] \right\}
 \nonumber \\
  & = & \varepsilon^2 \left\{  - \frac{ \beta ^4}{90}
   \delta ^2 (0) - \frac{ \beta ^3}{15}  \delta (0) \right\} .
\label{wbn@6***}\end{eqnarray}

 We now turn to the three-bubbles diagrams (\ref{wbn@f4}).
Among these,
there exist only three
involving the correlation function ${}_\mu \Delta _ \nu  (x,x')
\rightarrow  \dDeltad (\tau ,\tau ')$ for which
Eq.~(\ref{wbn@p7}) is not applicable:
the second, fourth, and sixth diagram.
 The other three-bubble
diagrams in (\ref{wbn@f4}) containing the
generalization
${}_\mu  \Delta _\mu (x,x)$
of the equal-time propagator
 $ \dDeltad (\tau ,\tau )$
can be calculated using Eq.~(\ref{wbn@p7}).

  Consider first a partial sum
consisting of the first,
third, and fifth three-bubble diagrams in the sum
(\ref{wbn@f4}).
This has the analytic form
\begin{eqnarray}
 \sum_{1,3,5} (\ref{wbn@f4}) & = &\frac{\varepsilon^2}{2}
  \int_0^\beta  \int_0^\beta  d\tau\,  d\tau ' \left\{ 4 \, \Delta  (\tau ,\tau
) \,
  \dDelta ^2 (\tau ,\tau ') \,\dDeltad (\tau ', \tau ') \right. \nonumber \\
 & + &  \left. 2\, \dDeltad (\tau ,\tau )  \Delta ^2 (\tau ,\tau ')
   \dDeltad (\tau ', \tau ') + 16 \,\dDelta
 (\tau ,\tau )  \Delta (\tau ,\tau ') \,\dDelta (\tau, \tau')\,
\dDeltad (\tau ',\tau ')
\right\} .
\label{wbn@7***}\end{eqnarray}
Replacing $\dDeltad (\tau ,\tau )$ and $\dDeltad (\tau ',\tau ')$
 by $ \delta (0) - 1/ \beta $
we see that Eq.~(\ref{wbn@7***}) contains,
with
opposite sign,
precisely the previous sum (\ref{wbn@32}) of all one-and two-bubble
diagrams. Together the give
\begin{eqnarray}
\!\!
 \sum (\ref{wbn@f3})
+\sum_{1,3,5}(\ref{wbn@f4})
& = & \frac{\varepsilon^2}{2 }
    \int_0^\beta  \int_0^\beta  d\tau\, d\tau ' \left\{ - \frac{4}{ \beta }
    \Delta (\tau , \tau ) \dDelta ^2 (\tau ,\tau ')
\right. \nonumber \\  & & \left.  ~~~~~~~~~~~~~~~~~~~~
 + \frac{2}{ \beta ^2}   \Delta ^2 (\tau ,\tau ') -
 \frac{16}{ \beta }  \,
 \dDelta  (\tau ,\tau )  \Delta (\tau ,\tau ')\, \dDelta
  (\tau ,\tau ') \right\} .
\label{wbn@8***}\end{eqnarray}
and can be evaluated directly to
\begin{eqnarray}
 \sum (\ref{wbn@f3})
+\sum_{1,3,5}(\ref{wbn@f4})
  & = & \frac{\varepsilon^2}{2} \left( - \frac{4}{ \beta }
\frac{ \beta ^2}{45} + \frac{2}{ \beta ^2}
	    \frac{ \beta ^4}{90} - \frac{16}{ \beta } \frac{ \beta ^3}{180}
 \right)
%\nonumber \\ & = &
= \frac{\varepsilon^2}{2} \left(-\frac{7}{45}
     \beta ^2 \right).
\label{wbn@9***}\end{eqnarray}
By the same direct calculation,
the Feynman integral in the  seventh three-bubble
diagram in (\ref{wbn@f4}) yields
\begin{eqnarray}
 I_7 & = & \int_0^\beta  \int_0^\beta  d\tau\, d\tau ' \,
\dDelta  (\tau ,\tau ) \dDelta (\tau ,\tau ')
\Deltad (\tau ,\tau ') \Deltad (\tau ',\tau ')
%\nonumber \\ & = &
= - \frac{ \beta ^2}{720}.
\label{wbn@10***}\end{eqnarray}
  The explicit results (\ref{wbn@9***})
and (\ref{wbn@10***}) are again not needed, since the last term in
Eq.~(\ref{wbn@8***}) is equal, with  opposite sign,
to the
partial sum of the sixth and seventh three-bubble diagrams in Eq.~(\ref{wbn@f4}).
To  see this, consider the Feynman integral associated
with the sixth three-bubble diagram in Eq.~(\ref{wbn@f4}):
\begin{eqnarray}
  I_6
 &= &
 \int_0^\beta  \int_0^\beta  d\tau\, d\tau '\, \dDelta
  (\tau ,\tau )  \Delta (\tau ,\tau ') \dDeltad
  (\tau ,\tau ') \Deltad (\tau ',\tau '),
\label{wbn@11***}\end{eqnarray}
whose
$d$-dimensional extension is
\begin{eqnarray}
  I_6^d
&= &
 \int_0^\beta  \int_0^\beta   d^dx\, d^dx'\, {}_\mu \Delta  (x,x)  \Delta
(x,x')
 {}_\mu  \Delta _ \nu  (x,x')  \Delta _ \nu (x',x').
\label{wbn@11***}\end{eqnarray}
Adding this to the seventh Feynman integral
(\ref{wbn@10***}) and performing a partial integration, we find
in one dimension
\begin{eqnarray}
 \sum_{6,7} (\ref{wbn@f4})  =  \frac{\varepsilon ^2}{2}\,16\,
\left(I_6 + I_7 \right)
 &=& \ve
\int_0^\beta  \int_0^\beta  d\tau\, d\tau ' \frac{16}{ \beta }
  \dDelta (\tau ,\tau ) \dDelta (\tau ,\tau ')  \Delta
 (\tau ,\tau ') \nonumber \\
 & = & \ve \left(\frac{4}{45}  \beta ^2 \right),
\label{wbn@12***}\end{eqnarray}
where we have used $d\left[ \dDelta (\tau ,\tau ) \right] /d\tau
 = -1/ \beta $ obtained by differentiating
(\ref{wbn@@2a***}).
Comparing
(\ref{wbn@12***})
with
(\ref{wbn@8***}),
we find the sum of all bubbles diagrams, except for the second
and fourth three-bubble diagrams in Eq.~(\ref{wbn@f4}), to be
 given by
\begin{equation}
  \sum(\ref{wbn@f3})+\sum_{2,4} {}^{'} (\ref{wbn@f4}) = \ve \left(-\frac{ \beta
^2}{15}\right).
\label{wbn@13***}\end{equation}
The prime on the sum denotes
 the exclusion of the diagrams
indicated  by subscripts.
 The correlation function
$\dDeltad (\tau , \tau ')$ in the
two remaining
diagrams of Eq.~(\ref{wbn@f4}),
whose $d$-dimensional extension
is
${}_\mu \Delta _ \nu (x,x') $,
cannot be replaced via Eq.~(\ref{wbn@p7}), and
the expression  can only be simplified
by applying
partial integration
 to the fourth diagram in
Eq.~(\ref{wbn@f4}), yielding
\begin{eqnarray}
 I_4 & = & \int_0^\beta  \int_0^\beta  d\tau\, d\tau '\,  \Delta (\tau ,\tau )
\dDelta
 (\tau ,\tau ') \dDeltad (\tau,\tau ') \Deltad (\tau ',\tau ')
\nonumber \\
& \rightarrow  & \int_0^\beta  \int_0^\beta  d^dx\, d^dx' \, \Delta (x,x)\,
{}_\mu \Delta
   (x,x') {}_\mu \Delta _ \nu  (x,x')  \Delta _ \nu
   (x',x') \nonumber \\
  & = &  \frac{1}2 \int_0^\beta  \int_0^\beta  d^dx\, d^dx' \, \Delta (x,x)
\Delta _  \nu
    (x',x') \partial'_ \nu  \left[ {}_\mu \Delta (x,x')\right]^2
 \nonumber \\
 & \rightarrow &
 \frac{1}2 \int_0^\beta  \int_0^\beta  d\tau\,d\tau '\, \Delta  (\tau ,\tau )
 \dDelta (\tau ',\tau ')
 \frac{d}{d\tau'} \left[ \dDelta^2 (\tau ,\tau ')\right]  \nonumber \\
 & = & \frac{1}{2\beta } \, \int_0^\beta  \int_0^\beta
    d\tau\, d\tau '  \Delta (\tau ,\tau ) \dDelta^2 (\tau ,\tau ') =
 \frac{\beta ^2}{90}.
\label{wbn@14***}\end{eqnarray}
The second diagram in the sum (\ref{wbn@f4}) diverges linearly.
As before, we add and subtract the divergence
\begin{eqnarray}
I_2 & = & \int_0^\beta  \int_0^\beta  d\tau\, d\tau ' \, \Delta (\tau ,\tau )
  \dDeltad^2 (\tau ,\tau ')  \Delta (\tau ',\tau ')
 \nonumber \\
 & = & \int_0^\beta  \int_0^\beta  d\tau\, d\tau '\, \Delta (\tau ,\tau )
  \left[ \dDeltad^2 (\tau ,\tau ') -  \delta ^2 (\tau -\tau ')
 \right]   \Delta (\tau ',\tau ')
 \nonumber \\ & + &
 \int_0^\beta  \int_0^\beta  d\tau\, d\tau '\,  \Delta ^2 (\tau ,\tau )
   \delta ^2 (\tau -\tau ').
\label{wbn@15***}\end{eqnarray}
In the first, finite  term
we go to $d$ dimensions
and replace $ \delta (\tau -\tau ') \rightarrow  \delta
 (x-x')=-
  \Delta _{\nu\nu} (x,x')$
using
the field equation (\ref{wbn@n5}).
After this, we apply  partial integration
and find
\begin{eqnarray}
 I^{R}_{2} & \rightarrow  &
 \int_0^\beta  \int_0^\beta  d^dx\, d^dx'\, \Delta (x,x) \left[ {}_\mu \Delta _
\nu ^2
  (x,x') -  \Delta _{ \lambda  \lambda } ^2 (x,x')
 \right]  \Delta (x',x')\nonumber \\
& = &\int_0^\beta  \int_0^\beta  d^dx\, d^dx'\,  \left\{ -\partial_\mu \left[
\Delta (x,x)
 \right]  \Delta_\nu (x,x')\,{}_\mu \Delta _ \nu (x,x')
  \Delta (x',x') \right.\nonumber \\
 &  &~~~~~~~~~~~~~~~~~~~~~~~~~+ \left. \Delta (x,x)  \Delta _ \nu  (x,x')
\Delta _{ \lambda  \lambda }
 (x,x') \partial_ \nu ' \left[  \Delta (x',x')\right] \right\}
 \nonumber \\
& \rightarrow &\int_0^\beta  \int_0^\beta  d\tau \,d\tau '\, 2 \left\{
 -\dDelta (\tau ,\tau )  \Deltad (\tau ,\tau ') \dDeltad (\tau ,\tau ')
   \Delta (\tau ',\tau ') + \right.\nonumber \\
 &  &~~~~~~~~~~~~~~~~~~~~~~~~~ \left.\Delta (\tau ,\tau )
\Deltad (\tau ,\tau ') \dDelta (\tau ',
\tau ')  \Deltadd (\tau ,\tau ')\right\} .
\label{wbn@16***}\end{eqnarray}
In going to the last line we have used $d[ \Delta (\tau ,\tau )]/d\tau  =
2\, \dDelta (\tau ,\tau )$ following from (\ref{wbn@@2a***}).
By interchanging
the order of integration $\tau  \leftrightarrow \tau '$,
the first term in Eq.~(\ref{wbn@16***})  reduced to the
integral (\ref{wbn@14***}).
In the  last  term
we replace $ \Deltadd (\tau ,\tau ')$
using  the field equation (\ref{wbn@eom}) and
 the trivial equation
\begin{equation}
 \int  d\tau \, \epsilon (\tau )\,  \delta (\tau ) = 0.
\label{wbn@37}\end{equation}
%
%as long as the singularity lies completely inside
%the integration region.
Thus we obtain
\begin{equation}
I_2=I_2^R+I_2^ {\div}
\label{wbn@n15***}\end{equation}
with
\begin{eqnarray}
 I^{R}_{2} & = & 2\left(  - \frac{ \beta ^2}{90}  -
  \frac{ \beta ^2}{120} \right)  = \frac{1}2 \left( - \frac{7 \beta ^2}{90}
    \right),\label{wbn@17***} \\
 I^{\div}_{2} & = &   \int_0^\beta \int_0^\beta  d\tau d\tau '
\Delta ^2 (\tau ,\tau )
  \delta ^2 (\tau -\tau ').
\label{wbn@18***}\end{eqnarray}
Using Eqs.~(\ref{wbn@14***}) and (\ref{wbn@n15***})
yields the sum of the second and fourth
three-bubble diagrams in Eq.~(\ref{wbn@f4}):
\begin{eqnarray}
 &&\!\!\!\!\!\!\!
\sum_{2,4} (\ref{wbn@f4}) = \frac{\varepsilon^2}{2}\,(2I_2 + 16I_4) =
%\nonumber\\&&
\varepsilon^2 \left\{ \int_0^\beta \int_0^\beta  d\tau
  d\tau '  \Delta ^2 (\tau ,\tau )  \delta ^2 (\tau -\tau ') +
\frac{  \beta  ^2}{20}\right\} .
\label{wbn@n19***}\end{eqnarray}
Finally, inserting this
into Eq.~(\ref{wbn@13***}), we
have the sum of all bubbles diagrams
\begin{equation}
  \sum(\ref{wbn@f4})+ \sum(\ref{wbn@f3}) = \varepsilon^2 \left\{ \int_0^\beta
\int_0^\beta  d\tau
  d\tau '  \Delta ^2 (\tau ,\tau )  \delta ^2 (\tau -\tau ') +
\frac{  \beta  ^2}{60}\right\} .
\label{wbn@19***}\end{equation}
Note that
 the finite part of this
is independent
of  ambiguous integrals of type (\ref{wbn@5a*}).

The contributions of the watermelon diagrams (\ref{wbn@f5}) correspond to the
Feynman integrals
\begin{eqnarray}
\!\!\!\!~\!\!\!\!\!\!\!\!\!\!\!\!\! \sum (\ref{wbn@f5}) &\!\! =\!\! & 2
\varepsilon^2 \int_0^\beta \int_0^\beta  d\tau d\tau ' \left[   \Delta ^2
       (\tau ,\tau ') \dDeltad^2 (\tau ,\tau ') \right.\nonumber \\
 &&~~~~~~~~~~~~ \left.~~~ + 4 \,\Delta  (\tau ,\tau ') \dDelta (\tau ,\tau ')
\Deltad (\tau ,\tau ')
   \dDeltad (\tau ,\tau ') + \dDelta^2 (\tau ,\tau ')  \Deltad^2
    (\tau ,\tau ') \right]     .
\label{wbn@40}\end{eqnarray}
The third integral is unique
and can be calculated
directly:
\begin{equation}
 I_{10} = \int^{ \beta }_{0} d\tau  \int^{ \beta }_{0}
     d\tau ' \,\dDelta^2 (\tau ,\tau ') \Deltad^2 (\tau ,\tau ')
 =  \varepsilon ^2\frac{ \beta ^2}{90}.
\label{wbn@20***}\end{equation}
The second integral reads in $d$ dimensions
\begin{eqnarray}
 I_9 & = & \int \int d^dxd^dx'  \Delta (x,x') {}_\mu \Delta (x,x')
       \Delta _ \nu  (x,x') \,{}_\mu \Delta _ \nu  (x,x').
\label{wbn@21***}\end{eqnarray}
This is integrated partially to yields, in one dimension,
\begin{equation}
 I_9 = - \frac{  1}{2} I_{10} - \frac{1}{2} \int\int d\tau\, d\tau '
    \Delta (\tau ,\tau ') \Deltad^2 (\tau ,\tau ')\, \ddDelta (\tau ,\tau ').
\label{wbn@22***}\end{equation}
 The  integral on the right-hand side is the one-dimensional version of
\begin{eqnarray}
I_{9'} & = &
\int_0^\beta \int_0^\beta  d^dxd^dx'  \Delta (x,x')  \Delta _ \nu ^2 (x,x') \,
\,     {}_{\mu\mu}  \Delta (x,x').
\label{wbn@23***}\end{eqnarray}
 Using the field equation
(\ref{wbn@n5}), going back to one dimension,
and inserting
 $ \Delta (\tau ,\tau' ), \Deltad (\tau ,\tau ')$,
and $\ddDelta (\tau ,\tau ')$
from
 (\ref{wbn@p4}), (\ref{wbn@p5}), and (\ref{wbn@eom}),
we perform all unique
integrals and
  obtain
\begin{equation}
   I_{9'} = - \beta ^2  \left\{ \frac{1}{24} \int d\tau  \,\epsilon^2
  (\tau )\,  \delta (\tau ) + \frac{1}{120} \right\} .
\label{wbn@24***}\end{equation}
Inserting this and (\ref{wbn@20***}) into Eq.~(\ref{wbn@22***})
gives, finally,
\begin{equation}
 I_9 = \left\{ \frac{1}{48} \int d\tau \,\epsilon ^2 (\tau )\,\delta (\tau )
  - \frac{1}{720} \right\}  \beta ^2.
\label{wbn@25***}\end{equation}

We now evaluate the
 first integral in
Eq.~(\ref{wbn@40}). Adding and subtracting
the linear divergence
 yields
%%%
%
\begin{eqnarray}
  I_8 &\!\! =\!\! & \int_0^\beta \int_0^\beta  d\tau \, d\tau '  \,\Delta ^2
(\tau ,\tau ')
  \dDeltad^2 (\tau ,\tau ') \nonumber \\
 & \!\!= \!\!& \int_0^\beta  \int_0^\beta  d\tau d\tau '  \Delta ^2 (\tau ,\tau
')
   \left[ \dDeltad^2 (\tau ,\tau ') -  \delta ^2 (\tau \!-\!\tau ') \right]
    + \varepsilon ^2 \int_0^\beta \int_0^\beta  d\tau  d\tau '  \Delta ^2 (\tau
,\tau )  \delta ^2
  (\tau\! -\!\tau ').
\label{wbn@26***}\end{eqnarray}
The finite part of the integral (\ref{wbn@26***}) has the
$d$-dimensional
extension
\begin{eqnarray}
 I^{R}_{8} & = &
 \int \int  d^dx\, d^dx'  \Delta ^2 (x,x') \left[
    {}_\mu \Delta _ \nu ^2 (x,x') -  \Delta ^2_{ \lambda  \lambda }
       (x,x') \right]
\label{wbn@27}\end{eqnarray}
which after partial integration
and going back
to one
dimension
reduces to a  combination of integrals Eqs.~(\ref{wbn@25***})
and (\ref{wbn@24***}):
\begin{equation}
 I^{R}_{8} = - 2 I_9 + 2 I_{9'} = - \left\{ \frac{1}{8}  \int
  d\tau \,\epsilon^2 (\tau )  \delta (\tau ) + \frac{1}{72}
 \right\}   \beta ^2 .
\label{wbn@28***}\end{equation}
The divergent part of $I_8$ coincides with
$I^{\div}_{2}$ in Eq.~(\ref{wbn@18***}):
\begin{equation}
 I^{\div}_{8} = \int_0^\beta  \int_0^\beta  d\tau d\tau '  \Delta ^2 (\tau
,\tau )
    \delta ^2 (\tau -\tau ')=I^{\div}_{2} .
\label{wbn@29***}\end{equation}
Inserting this together with
 (\ref{wbn@20***}) and (\ref{wbn@25***}) into Eq.~(\ref{wbn@40}), we obtain
the sum of watermelon
diagrams
\begin{eqnarray}
 \sum  (\ref{wbn@f5}) & = &  2\varepsilon^2 (I_8 + 4 I_9 + I_{10})
 \nonumber \\
 & = &  \varepsilon^2 \left\{ 2 \int_0^\beta \int_0^\beta  d\tau\, d\tau '
\Delta ^2
   (\tau ,\tau )  \delta ^2 (\tau -\tau ') - \frac{ \beta ^2}{12}
    \int_0^\beta  d\tau\, \epsilon^2 (\tau )  \delta (\tau ) - \frac{ \beta
^2}{60}
  \right\} .
\label{wbn@30***}\end{eqnarray}

For a flat space in curvilinear coordinates,
the sum of the first-order diagrams vanish.
To second order, the requirement of coordinate
independence
implies
the vanishing the sum of all connected diagrams
(\ref{wbn@f2})--(\ref{wbn@f5}). Setting the sum of Eqs.~(\ref{wbn@31}),
(\ref{wbn@19***}), and (\ref{wbn@30***}) to zero leads directly to the
integration  rule
(\ref{wbn@5a*}) and, in addition,
to the rule
\begin{equation}
 \int_0^\beta \int_0^\beta  d\tau\, d\tau '\,  \Delta ^2 (\tau ,\tau )  \delta
^2 (\tau -\tau ')
   =  \delta (0) \int d\tau \, \Delta ^2 (\tau ,\tau ),
\label{wbn@31***}\end{equation}
which we postulated  before in Eq.~(\ref{wbn@4*})
to cancel the $ \delta (0)$s coming from the measure
at the one-loop level.

The procedure can easily be continued to higher-loop diagrams
to define integrals over higher singular products of $\epsilon$-
and $ \delta $-functions. In this way we obtain
the confirmation of the
rule (\ref{wbn@4*}).
We have seen that at the one-loop level, the cancellation of $ \delta (0)$s
requires
\begin{equation}
 \int d\tau \, \Delta (\tau ,\tau )  \delta (0) =  \delta
 (0) \int d\tau \,  \Delta (\tau ,\tau ).
\label{wbn@32***}\end{equation}
The second-order equation~(\ref{wbn@31***})
contains the second power of
$\Delta (\tau ,\tau )$. To
$n$-order we find the equation
\begin{equation}
  \int d\tau _1 \dots d\tau _n
\Delta (\tau _1, \tau _2)
 \delta  (\tau _1, \tau _2)
      \cdots  \Delta  (\tau _n, \tau _1)
       \delta  (\tau _n, \tau _1)
=  \delta (0) \int d\tau\,  \Delta ^n (\tau ,\tau ).
\label{wbn@33***}\end{equation}
which reduces to
\begin{equation}
 \int \int d\tau _1 d\tau _n  \,\Delta ^n (\tau _1, \tau _1)
    \delta ^2 (\tau _1 - \tau _n)
=  \delta (0) \int d\tau  \,\Delta ^n (\tau ,\tau ),
\label{wbn@34***}\end{equation}
and this is satisfied given the rule
(\ref{wbn@4*}).  See Appendix C for a general derivation of these rules.

Let us now see what
the above integrals
imply for the
perturbation expansion
of the short-time amplitude
in curved space
in Riemann normal coordinates.
Taking into account the nonzeroth
contribution
(\ref{wbn@fosc}) of the first-order diagrams
reproduces immediately the first term
in the second-order operator expansion (\ref{wbn@sc}):
\begin{equation}
  \frac{1}{2} \langle {\cal A}_{\inter} \rangle ^2  =
 \frac{1}{2}\left(\varepsilon\,
  \frac{R}{12}\,\beta\right)^2 =
   \varepsilon ^2 \frac{R^2 }{288} \,\beta^2 \,.
\label{wbn@ft}\end{equation}
The sum of the local diagrams
(\ref{wbn@f2}) involves both tensors $R_{ij}^2$ and $R_{ijkl}^2$.
To order $\varepsilon^2$, we find
\begin{equation}
 \sum  (\ref{wbn@f2})  = -  \varepsilon ^2\frac{ \beta ^3}{30} \left(\frac{1}{36}
R^2_{ij}
 + \frac{1}{24} R^2_{ijkl} \right) \delta (0) + \varepsilon^2
 \frac{ \beta ^2}{24} \left(\frac{1}{45} R^2_{ij} +
  \frac{1}{30} R^2_{ijkl}\right).
\label{wbn@35***}\end{equation}
The contribution of
all bubbles  diagrams (\ref{wbn@f3}) and
(\ref{wbn@f4}) contains only $R_{ij}^2$:
\begin{equation}
 \sum  (\ref{wbn@f3}) + \sum (\ref{wbn@f4})  =
  \varepsilon ^2\frac{ \beta ^3}{1080} \,
 R^2_{ij}\, \delta (0) - \varepsilon^2
 \frac{ \beta ^2}{432} \, R^2_{ij} .
\label{wbn@n35***}\end{equation}
This compensates exactly the
$\delta (0)$-term proportional
to $R_{ij}^{2}$ in Eq.~(\ref{wbn@35***})
and yields correctly the
third second-order term $ - R_{ij}^{2}/720$
in the operator expansion (\ref{wbn@sc}).

Before turning to the contribution of
the second-order watermelon diagrams
(\ref{wbn@f5}) which contain initially
ambiguous Feynman integrals
we make an important observation.
Comparison with Eq.~(\ref{wbn@sc})
shows
that Eq.~(\ref{wbn@35***}) contains
already the correct part of the second-order
DeWitt-Seeley coefficient
$R_{ijkl}^2/720$.
Therefore, the only role of contributions
of the watermelon diagrams (\ref{wbn@f5}) which are proportional to
$R^2_{ijkl}$
must be to cancel a corresponding divergent part of the sum (\ref{wbn@35***}).
In fact,
the sum of the second-order watermelon diagrams (\ref{wbn@f5})
reads now,
\begin{equation}
 \sum (\ref{wbn@f5}) = \frac{ \varepsilon ^2}{24} R_{ijkl}^2  \left(I_8 - 2 I_9 +
I_{10}\right\} ,
\label{wbn@36***}\end{equation}
where the  integrals $I_8, I_9$, and $I_{10}$ were given before
in Eqs.~(\ref{wbn@29***}), (\ref{wbn@28***}), (\ref{wbn@25***}), and (\ref{wbn@20***}).
Substituting these into Eq.~(\ref{wbn@36***}) and using
the rules (\ref{wbn@4*}) and (\ref{wbn@5a*}), we obtain
\begin{equation}
 \sum (\ref{wbn@f5}) = \frac{ \varepsilon ^2}{24} R_{ijkl}^2 \int_0^\beta
\int_0^\beta  d\tau \,d\tau '
   \Delta ^2 (\tau ,\tau )  \delta ^2 (\tau -\tau ') =
   \varepsilon ^2 \frac{  \beta ^3}{720} R_{ijkl}^2 \, \delta (0),
\label{wbn@37***}\end{equation}
thus compensating the $\delta (0)$-term
proportional to $R_{ijkl}^{2}$ in
Eq.~(\ref{wbn@35***}) and no finite
contribution.

For one-component target space
as well as for $n$-component curved  space in normal coordinates,
our calculation procedure
using only the essence of
the $d$-dimensional extension
together
with the rules (\ref{wbn@4*}) and (\ref{wbn@5a*}) yields
unique  results
which
guarantee the coordinate independence
of path integrals and agrees
with
the
DeWitt-Seeley expansion
of the
short-time
amplitude.
The need for this
agreement
fixes the initially ambiguous integrals $I_8$ and $I_9$
to satisfy the equations
\begin{eqnarray}
 && I^{R}_{8} + 4 I_9 + I_{10} = - \frac{ \beta ^2}{120},\label{wbn@38***a}\\
 && I^{R}_{8} - 2 I_9 + I_{10} = 0,
\label{wbn@38***}\end{eqnarray}
as we can see from
 Eqs.~(\ref{wbn@30***}) and (\ref{wbn@36***}).
Since the integral
$I_{10}= \beta ^2/90$ is unique,
we must have $I_9 = - \beta ^2/720$ and
$I^{R}_{8} = - \beta ^2/72$, and this is what our integration rules
indeed gave us.

The main role
of the dimensional extension
in this context is to
forbid the application
of Eq.~(\ref{wbn@p7})
to
correlation
functions $\dDeltad (\tau ,\tau ')$.
This would have immediately fixed
 the finite part of the integral $I_8$
to the wrong value
 $I^{R}_{8} = - \beta ^2/18$,
leaving only
the integral $I_9$
which would define
 the
integral over distributions (\ref{wbn@5a*}).
In this way, however, we could only satisfy
 one of the equations
(\ref{wbn@38***a}) and
(\ref{wbn@38***}),
the other would always be violated.
 Thus, any regularization
different from ours  will ruin immediately
 coordinate independence.

It must be noted that
if we were to use arbitrary
 rather than Riemann normal coordinates,
one can
fix ambiguous integrals already at the two-loop level, and obtains the
conditions
(\ref{wbn@I12}).
Thus, although
the calculation in
 normal coordinates are simpler
and can be carried more easily to higher orders,
the perturbation in
arbitrary coordinates
help to fix more ambiguous integrals.

Let us finally compare our procedure
with the previous discussion of the same problem
by
 F.~Bastianelli,
P.~van Nieuwenhuizen,
and others
in Refs.~\cite{5,6,7,8,9,10,11,12,13}.
Those authors
suggested for almost ten years
two regularization schemes for perturbative
calculation on a finite-time interval: mode regularization
(MR) \cite{7,8,9} and time discretization (TS) \cite{9,10,11}.
They gave a detailed comparison of both schemes up to three loops
in Ref.~\cite{12}.
Their main goal was to
 calculate of trace anomalies of
quantum field theory by means of path integrals \cite{7,11,13}.
From the present point of view of extended
distribution theory,
mode regularization (MR) amounts to setting
\begin{equation}
\int d\tau  \,\epsilon^2 (\tau )\,\delta (\tau ) \equiv  \frac{1}{3}
{}.
\label{wbn@@}\end{equation}
%
%%%
With this rule,
the  ambiguous integrals $I_8$ and $I_9$
yield $I^{R}_{8} = - \beta ^2/18$, $I_9 =  \beta ^2/180$.
However, these values do not allow
for
coordinate independence, nor do they lead to the
correct short-time
DeWitt-Seeley expansion of the amplitudes. This is
what forced the authors
 to add an unpleasant
noncovariant ``correction term"
${\cal A}^{\rm fudge} = -\int d\tau   \Gamma ^i_{jk}  \Gamma ^l_{mn}
  g_{il}g^{im}g^{kn}/24
$
to the classical action,
in violation of Feynman's construction rules for path integrals.
In doing this they followed earlier work
by Salomonson in Ref.~\cite{6}.

Their time discretization scheme (TS), on the other hand,
amounts to setting
\begin{equation}
\int d\tau \epsilon^2 (\tau )\,\delta (\tau ) = 0.
\label{wbn@@zero}\end{equation}
They applied this to purely one-dimensional
calculations which,
as we have shown in this paper,
 leads to the contradictory results
 depending  on
where partial integration or field equations are used.
While $I_8$ is again $I^{R}_{8} = -  \beta ^2/18$,
the result for $I_9 = 7  \beta ^2/360$ is not unique.
To obtain  coordinate independence as well as the correct
DeWitt-Seeley expansion, they had now to add another
noncovariant ``correction term"
${\cal A}^{\rm fudge} =\int d\tau   \Gamma _{jk}^i  \Gamma _{il}^j g^{kl}/8$,
thereby
following
the original work of Gervais and Jevicki in Ref.~\cite{5}.

In recent papers \cite{14,15,BD}, the authors
of Refs.~\cite{8}
and \cite{12}
have
begun following
our method of dimensional regularization
 \cite{2,3}, adapting it to
a finite time interval in  \cite{15}.
They now obtain,
of course, correct
coordinate-independent results
without noncovariant additional terms in the action.
They do not, however,  exhibit
the precise
location of ambiguities
as we did here, and most importantly,
they do not
derive from their results
rules
for  integrating
products of $ \epsilon $- and $ \delta $-functions,
which are central to the present paper.
In particular, they do not
realize that
dimensional regularization
amounts to the integration rule Eq.~(\ref{wbn@@zero}).

{}~\\
Acknowledgment:\\
This work was financed  in part by Deutsche Forschungsgemeinschaft
under Grant  Kl 256-22.

\vspace{1.5cm}

\section{Appendix A: Integrals $I_{14}$ and $I_{15}^{R}$
from two-loop expansion in arbitrary coordinates}~~

To order $\varepsilon$, the metric and its determinant have the
expansions:
\begin{eqnarray}\label {a1}
g_{ij}(q) &=& \delta_{ij} + \sqrt{\varepsilon}(\partial_{k} g_{ij})q^{k} +
\varepsilon\frac{1}{2} (\partial_l
\partial_k g_{ij}) q^l q^k,\nonumber\\
\log g(q) &=& \sqrt{\varepsilon} g^{ij} (\partial_k
g_{ij})q^k + \varepsilon \frac{1}{2} \,g^{ij}\,[ (\partial_l
\partial_k g_{ij}) - g^{mn}(\partial_l
g_{im})(\partial_k g_{jn})] q^l q^k.
\end{eqnarray}
The interaction (\ref{wbn@6**}) becomes
\begin{eqnarray}\label{wbn@a2}
\!\!\!\!\!\!\!\!\!\!\!{\cal A}_{\inter} [q] &=&
\int^{\beta}_{0} d\tau \Big\{\Big[\frac{1}{2}\sqrt{\varepsilon}\,
(\partial_k g_{ij}) q^k + \frac{1}{4}\varepsilon\, (\partial_l
\partial _k g_{ij}) q^l q^k \Big] \dot q^i \dot q^j \nonumber\\
&& -\frac{1}{2}\sqrt{\varepsilon}\, \delta(0) g^{ij} (\partial_k g_{ij})
q^k
- \frac{1}{4} \varepsilon\, \delta (0)\, g^{ij} \Big[
(\partial_l \partial_k g_{ij}) - g^{mn}\,(\partial_l g_{im})(\partial_k
g_{jn})\Big] q^l q^k \Big\}.
\end{eqnarray}

To the first-order in $\varepsilon$,
the perturbation expansion (\ref{wbn@7**})
with the interaction (\ref{wbn@a2})
consist of two sets of diagrams
proportional to $\Gamma_{ij,\,k}$
and $\Gamma^{2}_{ij,\,k}$, respectively.
First, there are the same local diagrams as in Eq.~(\ref{wbn@f1}):
the first two local diagrams
coming from the kinetic part
of
 (\ref{wbn@a2})
carry a factor $\Gamma_{ij,\,k}$,
while the last local diagram, coming from
the measure part of (\ref{wbn@a2}),
involves
both factors $\Gamma_{ij,\,k}$
and  $\Gamma^{2}_{ij,\,k}$.
Omitting the $\Gamma^{2}_{ij,\,k}$-part
of the last diagram,
the terms linear in the
Christoffel symbol
$\Gamma_{ij,\,k}$
coming from the sum of
local diagrams in (\ref{wbn@f1})
reads
\begin{eqnarray}\label{wbn@a3}
\!\!\sum (\ref{wbn@f1}) &=& -{\varepsilon\over 4} (\partial_l \partial_k
g_{ij})\int^\beta_0 d\tau \Big\{ g^{ij} g^{kl}
\dDeltad (\tau,\tau) \Delta (\tau,\tau) +
%\nonumber\\&& \,\,\,\,\,\,\quad\quad +\,
2 g^{ik} g^{jl}\,\,
\dDelta ^2 (\tau,\tau)\! -\! \delta (0) g^{ij} g^{kl}
\Delta (\tau,\tau) \Big\}
\nonumber\\
&=&\beta\frac{\varepsilon}{24}
(\partial_l \partial_k g_{ij})(g^{ij}g^{kl} - g^{ik}g^{jl})
%\nonumber\\&&\,\,\,\,\,\,\quad\quad = \,
=\beta\frac{\varepsilon}{24}
g^{ij} g^{kl} (\partial_l \Gamma_{ik,\,j} - \partial_i \Gamma_{lk,\,j}).
\end{eqnarray}
In addition,
the interaction (\ref{wbn@a2}) generates
nonlocal
first-order diagrams proportional to $\Gamma_{ij,\,k}^2$.
Together with nonlinear in Christoffel symbol
part of the last local diagram in Eq.~(\ref{wbn@f1}),
they are represented as follows
\begin{eqnarray}\label{wbn@a4}
&&~~\,{\varepsilon\over 2 } g^{lk}\, \Gamma_{li}{}^i \,\Gamma_{kj}{}^j
\big[
\hspace{0mm}\raisebox{-1mm}{\mbox{\input ele.tps }} -\, 2\delta (0)\hspace{0mm}\raisebox{-1mm}{\mbox{\input el.tps }} + \,\,
\delta^2 (0)\!\!\!\hspace{0mm}\raisebox{-1mm}{\mbox{\input q.tps }}
\big]
\nonumber\\
&&+\, \varepsilon \Gamma_{li}{}^i\,(g^{lk}\,\Gamma_{kj}{}^j
+ g^{jk}\,\Gamma_{jk}{}^l )
\big[
\hspace{0mm}\raisebox{-1mm}{\mbox{\input hhe.tps }} -\, \delta (0)\hspace{0mm}\raisebox{-1mm}{\mbox{\input hh.tps }}
\big]
\nonumber\\
&&+\,{\varepsilon\over 2}
 (g^{il}\,g^{kn}\,\Gamma_{il}{}^j\,\Gamma_{kn,\,j} + g^{ij}\,
\Gamma_{ik}{}^k\,\Gamma_{jl}{}^l + 2g^{ij}\,\Gamma_{ij}{}^k\,
\Gamma_{kl}{}^l)
\hspace{0mm}\raisebox{-1mm}{\mbox{\input hoh.tps }}
\nonumber\\
&&+\,{\varepsilon\over 2}  (g^{ik}\,
g^{jl}\,\Gamma_{il}{}^n \,\Gamma_{kj,\,n} + 3g^{ik}\,
\Gamma_{il}{}^n\, \Gamma_{nk}{}^l)
\hspace{0mm}\raisebox{-1mm}{\mbox{\input the112.tps }}
\nonumber\\
&&+\,{\varepsilon\over
 2}  g^{lk}\,(\Gamma_{lj}{}^i\,\Gamma_{ik}{}^j + g^{in}\,
\Gamma_{nk}{}^j\, \Gamma_{il,\,j})
\big[\hspace{0mm}\raisebox{-1mm}{\mbox{\input the022.tps }} -\, \delta (0) \hspace{0mm}\raisebox{-1mm}{\mbox{\input 0dot.tps }}\big]
\end{eqnarray}
\comment{ Finally, there is one-loop first-order diagram coming from the
measure in Eq.~(\ref{wbn@a2}):
\begin{equation}\label{wbn@a5}
{\varepsilon\over 8} 4g^{lk} \Gamma^i_{li} \Gamma^j_{kj}
\delta^2 (0) - {\varepsilon\over 8}\, 4\, \big[ g^{ij} g^{lk}
\Gamma_{ik,\,n} \Gamma^n_{jl} + g^{lk} \Gamma^n_{ki}
\Gamma^i_{ln} \big] \delta (0) \hspace{0mm}\raisebox{-1mm}{\mbox{\input 0dot.tps }} .
\end{equation}
}
The Feynman integrals associated with the diagrams
in the first and second lines of Eq.~(\ref{wbn@a4}) read
\begin{eqnarray}\label{wbn@a6}
\!\!\!\!\!\!\!
\!\!\!\!\!\!\!
I_{11} &=& \int\int d\tau\, d\tau' \,\left\{ \dDeltad
(\tau,\tau)
\Delta (\tau,\tau') \dDeltad (\tau',\tau')
%\right.\nonumber\\&-& \left.
-2\delta (0)\,\dDeltad (\tau, \tau)\, \Delta (\tau, \tau')
+ \delta^2 (0)\,\Delta (\tau, \tau')\right\}
\end{eqnarray}
and
\begin{equation}\label{wbn@a8}
  I_{12} = \int\int d\tau\, d\tau' \,\left\{
\dDelta (\tau, \tau) \dDelta
  (\tau,\tau') \dDeltad (\tau',\tau') - \delta (0)\,\dDelta (\tau, \tau)
\dDelta (\tau, \tau')\right\},
\end{equation}
respectively.
Replacing in Eqs.~(\ref{wbn@a6}) and (\ref{wbn@a8})
$\dDeltad (\tau,\tau)$ and
$\dDeltad (\tau',\tau')$ by
$\delta(0) - 1/\beta$ leads to cancellation
of the infinite factors $\delta (0)$
and $\delta^2 (0)$ coming from the measure,
such that we are left with
\begin{equation}\label{wbn@a7}
I_{11} =
 {1\over{\beta^2}}\, \int_0^\beta d\tau \int^\beta_0 d\tau'
\Delta (\tau,\tau') = {\beta \over 12}
\end{equation}
and
\begin{equation}\label{wbn@a9}
  I_{12} = - {1\over\beta} \,\int^\beta_0 d \tau
  \int^\beta_0 d\tau' \,\dDelta (\tau,\tau)
  \dDelta (\tau,\tau')= -
  {\beta \over 12}.
\end{equation}

The Feynman integral
of the diagram
in the third line
of Eq.~(\ref{wbn@a4}) has $d$-dimensional extension
\begin{eqnarray}&&\!\!\!\!\!  I_{13} =\int\int d\tau\, d\tau' \,\dDelta (\tau,\tau)
  \Deltad (\tau',\tau') \dDeltad (\tau,\tau')
% \nonumber\\&&
 \to \int\int d^dx\, d^dx' {_\mu}\Delta (x,x)
  \Delta_{\nu} (x',x') {_\mu}\Delta_{\nu} (x,x').
\nonumber \\&&
\label{wbn@a10}
\end{eqnarray}
Integrating this partially yields
\begin{eqnarray}\label{wbn@a11}
  && \!\!\!\!\!\!\!I_{13} = {1\over\beta} \int\int d\tau\, d\tau'
  \Deltad (\tau,\tau') \dDelta
  (\tau',\tau')
%= \nonumber\\  &&
= {1\over\beta} \int^\beta_0 d\tau \int^\beta_0
  d\tau'\, \dDelta (\tau,\tau) \dDelta
  (\tau,\tau') = {\beta\over 12},
\end{eqnarray}
where we have
interchanged
the order of integration $\tau  \leftrightarrow \tau '$
in the second line of Eq.~(\ref{wbn@a11})
and used $d [\dDelta (\tau,\tau)] /d\tau = - 1/\beta.$
Multiplying the integrals (\ref{wbn@a7}),
(\ref{wbn@a9}), and (\ref{wbn@a11}) by corresponding
vertices in Eq.~(\ref{wbn@a4}) and adding  them together,
we obtain
\begin{equation}\label{wbn@a12}
  \sum_{1,2,3} (\ref{wbn@a4}) =
 \frac{\varepsilon\beta}{24}\, g^{ij} g^{kl}\,
  \Gamma_{ij}{}^n\, \Gamma_{kl,\,n} .
\end{equation}

The contributions of the
last three diagrams
in the fourth and the fifth
line of Eq.~(\ref{wbn@a4})
correspond to the
ambiguous
integrals (\ref{wbn@7*}) and (\ref{wbn@8*}),
respectively.
Moreover,
the difference of two diagrams
in the last line of Eq.~(\ref{wbn@a4})
contains only the finite part
of the integral (\ref{wbn@8*}),
since
its
divergent part (\ref{wbn@13*})
is canceled by the contribution of the local diagram
with the factor $\delta (0)$.
Multiplying these integrals
by corresponding vertices
in Eq.~(\ref{wbn@a4})
yields
the sum of diagrams
in the fourth and the fifth line
of Eq.~(\ref{wbn@a4}) %generally
as follows
\begin{eqnarray}\label{wbn@a13}
 &&\!\!\!\!\!\! \sum_{4,5} (\ref{wbn@a4}) =
 \frac{\varepsilon}{2}\,\Big\{ g^{ik}\, g^{jl}\,
  \Gamma_{il}{}^n\,\Gamma_{kj,\,n}\,\left(I_{14} +
 I^{R}_{15}\right)
%\nonumber\\&&\quad\quad
+ \, g^{lk}\,\Gamma_{lj}{}^i\,\Gamma_{ik}{}^j\,
 \left(3I_{14} + I^{R}_{15}\right)
\Big\} .
\end{eqnarray}

On the other hand,
to guarantee
the coordinate independence
of path integrals,
this sum must be
\begin{equation}\label{wbn@a14}
  \sum_{4,5} (\ref{wbn@a4}) = -  \frac{\varepsilon\beta}{24}\,
 g^{ij} g^{kl}\,
  \Gamma_{ik}{}^n\,
  \Gamma_{jl,\,n} .
\end{equation}
Adding this
to
(\ref{wbn@a12}),
we find the sum of all diagrams
in (\ref{wbn@a4}) as follows
\begin{equation}\label{wbn@a15}
  \sum (\ref{wbn@a4}) = \frac{\varepsilon\beta}{24}\, g^{ij} g^{kl}\,\big(
  \Gamma_{ij}{}^n \Gamma_{kl,\,n} - \Gamma_{ik}{}^n
  \Gamma_{jl,\,n}).
\end{equation}
Together with the sum over all diagrams in
(\ref{wbn@f1}) calculated in  (\ref{wbn@a3})
this yields,
finally, the sum of all first-order
diagrams
\begin{equation}\label{wbn@a16}
\sum (\ref{wbn@f1}) + \sum (\ref{wbn@a4}) =
{\varepsilon\beta\over 24}\, g^{ij} g^{kl}\,
R_{likj} = -{\varepsilon\beta\over 24}\, R.
\end{equation}
The result is perfectly covariant and agrees, of course,
 with Eq.~(\ref{wbn@3***})
derived in normal coordinate.
Comparing now Eq.~(\ref{wbn@a13}) with (\ref{wbn@a14}),
we find
\begin{eqnarray}\label{wbn@a17}
 I_{14} + I^{R}_{15} &=& - \frac{\beta}{12}\,, \nonumber\\
 3I_{14} + I^{R}_{15} &=& 0\,.
\end{eqnarray}
Thus, coordinate independence
specifies the
initially ambiguous
integrals (\ref{wbn@7*}) and (\ref{wbn@8*})
to have indeed the values (\ref{wbn@I12}).

\section{Appendix B: Operator derivation of short-time
DeWitt-Seeley expansion}
Here we give a short derivation of the
DeWitt-Seeley expansion
(\ref{wbn@b1a}).
 In a neighborhood of
some arbitrary point $q_0^i$ we expand
the
Laplace-Beltrami operator in
normal coordinate system (\ref{wbn@12**}) as
\begin{equation}\label{wbn@b2}
  D^2=\partial^2 - {1\over 3} R_{ik_1 jk_2}(q_0) (q - q_0)^{k_1}
  (q - q_0)^{k_2} \partial_i \partial_j - {2\over 3} R_{ij}(q_0)
  (q - q_0)^i \partial_j.
\end{equation}

To find the coefficients $a_k(q,q')$ in Eq.~(\ref{wbn@b1a}), we resort to
perturbation theory. The time displacement operator $H=-D^2/2$
in the exponent
of Eq.~(\ref{wbn@b1a}) is
separated into a free part $H_0$ and an interaction part
$H_{\inter}$ as follows
\begin{eqnarray} H_0 &=& - {1\over 2} \partial^2,\label{wbn@b3a} \\
 H_{\inter} &=& {1\over 6} R_{ik_1 jk_2} (q-q_0)^{k_1} (q-q_0)^{k_2}
 \partial_i \partial_j + {1\over 3} R_{ij} (q-q_0)^i \partial _j.
\label{wbn@b3}
\end{eqnarray}
The  transition amplitude (\ref{wbn@b1a})
satisfies the integral equation
\begin{eqnarray}\label{wbn@b4}
 \!\!\!\!\!\!
(q,\beta\mid q',0)
& =& \langle q\mid e^{-\beta (H_0+H_{\inter})} \mid q'\rangle
=  \langle q\mid e^{-\beta H_0}\left[1-\int_0^ \beta \!
 d \sigma e^{ \sigma  H_0} H_{\inter} e^{- \sigma  H}\right]
 \mid q'\rangle
\nonumber \\&=&
(q,\beta\mid q',0)_0
- \int^\beta_0\!\!d\sigma\! \int\!\! d^n \bar q \,
(q,\beta- \sigma \mid \bar q,0)_0\, H_{\inter} (\bar q) \,
(\bar{q}, \sigma \mid q,0),
\end{eqnarray}
where
\begin{equation}\label{wbn@b5}
  (q,\beta\mid q',0)_0=
 \langle q\mid e^{-\beta H_0} \mid q'\rangle=
\frac{1}{ \sqrt{2\pi  \beta}^{n}}\,
   e^{- {(\Delta q)^2}/{2\beta}}.
\end{equation}
To  first order in
$H_{\inter}$ we obtain
\begin{eqnarray}\label{wbn@b6}
(q,\beta\mid q',0)
=(q,\beta\mid q',0)_0
- \int^\beta_0\!\!d\sigma\! \int\!\! d^n \bar q \,
\,(q,\beta- \sigma \mid \bar q,0)_0\, H_{\inter} (\bar q) \,
(\bar{q}, \sigma \mid q,0)_0.
\end{eqnarray}
Inserting (\ref{wbn@b3}) and choosing $q_0=q'$, we find
\begin{eqnarray}\label{wbn@b7}
(q,\beta\mid q',0)
&=&(q,\beta\mid q',0)_0
 \left\{ 1+ \int^\beta_0
d\sigma \int \frac{d^n (\Delta \bar q)}{ \sqrt{2\pi a}^{n}}\,
e^{-[\Delta \bar q - (\sigma/\beta)\,\Delta q ]^{2}/2a} \right.
\nonumber\\
&\times &  \left.
\left[ - {1\over 6} R_{ik_1jk_2}
\Delta \bar q^{k_1} \Delta \bar q^{k_2} \left( -
{\delta^{ij} \over \sigma} + {\Delta \bar q^i
\Delta \bar q^j\over \sigma^2}\right) + {1\over 3} R_{ij}
\frac{\Delta \bar q^i \Delta \bar q^j}{\sigma}
\right]\right\},
\end{eqnarray}
where we have replaced the integrating variable $\bar q$ by
$\Delta \bar q = \bar q - q'$ and used the notation $a=
(\beta-\sigma)\sigma/\beta$. There is initially also a term of fourth order
in $\Delta \bar q$ which vanishes, however,
 because of the antisymmetry of
$R_{i k j l}$ in $ik$ and $jl$.
The  remaining
Gaussian integrals
are performed after
shifting $\Delta \bar q\to \Delta \bar q +
\sigma\,\Delta q/ \beta $,
 and    we obtain
\begin{eqnarray}\label{wbn@b8}
(q,\beta\mid q',0)
&=&
(q,\beta\mid q',0)_0
 \Big\{ 1+{1\over 6}
\int^\beta_0 d\sigma \Big[ {\sigma\over\beta^2} R_{ij}
(q')\Delta q^i \Delta q^j +
{a\over \sigma} R(q') \Big]\Big\}\nonumber \\
& =&
(q,\beta\mid q',0)_0
 \Big[ 1 +
{1\over 12} R_{ij} (q') \Delta q^i \Delta q^j +
{ \beta\over 12}  R(q') \Big].
\end{eqnarray}
Note that all geometrical quantities are evaluated at the initial point
$q'$. They can be re-expressed in power series around the final
position $q$ using the fact that  in normal coordinates
\begin{eqnarray}&& g_{ij} (q') = g_{ij} (q) + {1\over 3} R_{i k_1 j k_2} (q)
\Delta q^{k_1} \Delta q^{k_2}+\dots~, \label{wbn@b9a}\\
&&
g_{ij} (q') \Delta q^i \Delta
q^j = g_{ij} (q) \Delta q^i \Delta q^j,
\label{wbn@b99}
 \end{eqnarray}
the latter equation being true to all orders in $ \Delta q$
due to the antisymmetry of the tensors $R_{ijkl}$ in all terms of
the expansion (\ref{wbn@b9a}), which
is just another form of writing the expansion (\ref{wbn@12**a})
up to the second order in $ \Delta q^i$.

Going back to the general coordinates, we
  obtain all
 coefficients of the expansion (\ref{wbn@b1a})  linear in
the curvature tensor
\begin{equation}\label{wbn@b10}
(q,\beta\mid q',0)
 \simeq
\frac{1}{ \sqrt{2\pi  \beta}^{n}}\,
 e^{-
g_{ij} (q) \Delta q^i \Delta q^j/2 \beta } \Big[ 1+
{1\over 12} R_{ij} (q)\Delta q^i \Delta q^i +
{\beta\over 12} R(q)\Big].
\end{equation}
 The higher terms in (\ref{wbn@b1a})
can be derived similarly, although with much more effort.

A simple cross check of the expansion
(\ref{wbn@b1a})
to high orders
is possible
if we restrict the space to a sphere
of radius $r$
in $D$ dimensions. Then
\begin{equation}\label{wbn@d1}
 R_{ijkl} = - \frac{1}{r^2}\, \left( g_{ik}\,g_{jl} -
 g_{il}\,g_{jk}\right), \,
~~~ i,j = 1,2,\dots,n = D-1,
\end{equation}
where $n = 2$ is dimension of a sphere,
and $D = 3$ is dimension of a flat
embedding space, respectively.
Contractions yield Ricci tensor and scalar curvature
\begin{eqnarray}\label{wbn@d2}
 R_{ij} &=& R_{kij}{}^k = \frac{D-2}{r^2}\,g_{ij}
 ,~~~~~~~
 R = R_{i}{}^i = \frac{(D-1)(D-2)}{r^2}
\end{eqnarray}
and further:
\begin{eqnarray}\label{wbn@d3}
 R_{ijkl}^{2} = \frac{2(D-1)(D-2)}{r^4} ,~~~~~
 R_{ij}^{2} = \frac{(D-1)(D-2)^2}{r^4} .
\end{eqnarray}
Inserting these into (\ref{wbn@sc}),
we obtain
the DeWitt-Seeley short-time
expansion of the amplitude
from $q=0$ to
 $q=0$ up to order $ \beta ^2$:
\begin{eqnarray}&&\!\!\!\!\!\!\!(0,\beta\mid 0,0) =
\frac{1}{ \sqrt{  2\pi  \beta}^{D\!-\!1}}\,
 \left[ 1 +
 (D\!-\!1)(D\!-\!2)\frac{\beta}{12r^2} + (D\!-\!1)(D\!-\!2)(5D^2-17D+18)
\frac{\beta^2 }{ 1440r^4}
  \right].                   \nonumber \\&&
\label{wbn@d4}
\end{eqnarray}

On the other hand, we may follow Ref.~\cite{MOD},
and  calculate explicitly
the partition function
for this system
\begin{equation}\label{wbn@d6g}
\!\!\!\!\!\! Z (\beta) = \sum_{l=0}^{\infty}\,
d_l\,\exp[-l(l+D\!-\!2)\beta/2r^2]\,\,,
\end{equation}
where $-l(l+D-2)$ are the eigenvalues
of the Laplace-Beltrami operator on a sphere
and $d_l=(2l+D-2)(l+D-3)!/l!(D-2)!$  their degeneracies.
Since the space is homogeneous, the
amplitude   $(0,\beta\mid 0,0)$ is  obtained from this
by dividing  out the
 constant surface
of a sphere:
\begin{equation}
     (0,\beta\mid 0,0)=\frac{\Gamma(D/2)}{2\pi^{D/2}r^{D-1}}Z( \beta ).
\label{wbe@Divout}\end{equation}

For any given $D$, the sum
in (\ref{wbn@d6g}) easily be expanded in powers of
$ \beta $.
As an example, take $D=3$ where
\begin{equation}\label{wbn@d6}
\!\!\!\!\!\! Z (\beta) = \sum_{l=0}^{\infty}\,
  (2l+1)\,\exp[-l(l+1)\beta/2r^2]\,\,.
\end{equation}

In the small-$\beta$ limit,
the sum (\ref{wbn@d6})
is  evaluated
as follows
\begin{eqnarray}\label{wbn@d7}
\!\!\!\!\!\!Z (\beta) =
 \int_{0}^{\infty}\,d\,[l(l+1)] \, \exp [-l(l+1)\beta/2r^2] +
 \sum_{l=0}^{\infty}\,(2l+1)\left[ 1 -
  l(l+1)\beta /2r^2 + \,\dots\,
  \right]\,.
\end{eqnarray}
The integral is immediately done
and yields
\begin{eqnarray}\label{wbn@d8}
&& \int_{0}^{\infty}\,d z \, \exp (-z\beta/2r^2) =
 \frac{2r^2}{\beta}\,.
\end{eqnarray}
The sums are divergent but can be evaluated by analytic continuation
from negative powers of $l$ to positive ones with the help
of Riemann zeta functions $\zeta(z)=\sum_{n=1}^\infty n^{-z}$,
which vanishes for all even negative arguments.
Thus we find
\begin{eqnarray} \!\!\!\!\!\!\!\!\!\!\sum_{l=0}^{\infty}\,(2l+1) &=& 1 +
  \sum_{l=1}^{\infty}\,(2l+1) = 1 + 2\zeta (-1) - \frac{1}{2}  =
    \frac{1}{3}\,,\label{wbn@dd1}\\
\!\!\!\!\!\!\!\!\!\!-  \frac{\beta}{2r^2}\,\sum_{l=0}^{\infty}\,(2l+1)l(l+1) &=& -
  \frac{\beta}{2r^2}\,\sum_{l=1}^{\infty}\,(2l^3 + l) =-
  \frac{\beta}{2r^2}\,[2\zeta (-3) + \zeta (-1)] =
   \frac{\beta}{30r^2}\,.
\label{wbn@dd}
 \end{eqnarray}
Substituting these into (\ref{wbn@d7}),
we find
\begin{eqnarray}\label{wbn@d5}
\!\!\!\!\!\!\!\!\!\!Z (\beta) &=&
\frac{2r^2}{\beta}\,
 \left( 1 +
 \frac{\beta}{6r^2} + \frac{\beta^2 }{ 60 r^4}
  +\dots\right).
\end{eqnarray}
Dividing
  out the
 constant surface
of a sphere $4\pi r^2 $ as required by Eq.~(\ref{wbe@Divout}), we obtain
indeed
the expansion (\ref{wbn@d4})
  for $D=3$.

\section{Appendix C: Cancellation of all powers of $ \delta (0)$}
 There is a simple way of proving
the cancellation of all UV-divergences $ \delta (0)$.
Consider
a free particle whose mass depends on the time
with an action
\begin{equation}
 {\cal A}_{\rm tot} [q] = \int^{ \beta }_{0} d\tau
   \left[ \frac{1}{2} Z (\tau ) \dot q^2 (\tau ) - \frac{1}{2}
   \delta (0) \log Z (\tau )\right] ,
\label{wbn@c1}\end{equation}
where $Z (\tau)$ is some function of $\tau$ but
independent now of the path $q (\tau)$.
The last term is the simplest
nontrivial form of the Jacobian action
in (\ref{wbn@4**}). Since it is independent of $q$,
it is conveniently taken out of the path integral as
a factor
\begin{equation}
  J = e^{(1/2 ) \delta (0) \int_{0}^{ \beta } d\tau  \log
  Z (\tau )}.
\label{wbn@c3}\end{equation}
We split the action into a sum of a  free and an interacting part
\begin{eqnarray}
 {\cal A}_0= \int_{0}^{ \beta } d\tau   \,
 \frac{1}{2}  \dot q^2 (\tau ),~~~~
 {\cal A}_{\rm int}
= \int_{0}^{ \beta } d\tau   \,
 \frac{1}{2}\left[Z(\tau )-1\right]  \dot q^2 (\tau ),~~~~
\label{wbn@@}\end{eqnarray}
and
calculate
the transition amplitude (\ref{wbn@7**})
as a sum of all connected diagrams
in the cumulant expansion
\begin{eqnarray}
 \langle 0,  \beta  \vert 0,0\rangle & = & J \, \int {\cal D} q (\tau )
     e^{-{\cal A}_{0}  [q] - {\cal A}_{\inter} [q] }
% \nonumber \\ & = &
= J \, \int {\cal D}  q (\tau ) e ^{-{\cal A}_{0} [q]} \left( 1 -
    {\cal A}_{\inter} + \frac{1}{2} {\cal  A}^2_{\inter} -
\dots \right)
 \nonumber \\
 & = &  (2 \pi  \beta )^{-1/2}J \, \left[ 1 - \langle {\cal A}_{\inter}
          \rangle + \frac{1}{2} \langle {\cal A}^2_{\inter} \rangle -
    \dots \right]
\nonumber \\
 & = &  (2 \pi  \beta )^{-1/2}J \, e^{ - \langle {\cal A}_{\inter}
          \rangle_c + \frac{1}{2} \langle {\cal A}^2_{\inter} \rangle_c -
    \dots } .
\label{wbn@c2}\end{eqnarray}
We now show that
the
 infinite series the of $\delta (0)$-powers
appearing in
a  Taylor expansion of the exponential
(\ref{wbn@c3})
is precisely compensated by the sum of
all terms in the perturbation
expansion (\ref{wbn@c2}).
Being interested only in these singular terms,
we may extend the $\tau $-interval
to the entire time axis. Then
Eq.~(\ref{wbn@p7}) yields the propagator
$\dDeltad(\tau ,\tau ')= \delta (\tau -\tau ')$,
 and we find
the  first-order expansion term
\begin{equation}
\langle {\cal A}_{\inter}
          \rangle_c
 = \int d\tau \, \frac{1}{2} [Z(\tau )-1]\,\dDeltad (\tau ,\tau )   = -\, \frac{1}{2}
\delta (0) \int d\tau \, [1-Z(\tau )].
\label{wbn@c4}\end{equation}
To second order,
divergent integrals appear involving products of distributions,
thus  requiring
an intermediate
extension to $d$ dimensions
 as follows
\begin{eqnarray}
\langle {\cal A}^2_{\inter} \rangle_c
 &=& \int\int d\tau_1\, d\tau_2 \, \frac{1}{2}(Z-1)_1 \,\frac{1}{2}(Z-1)_2\,
 2\,\dDeltad  (\tau_1 ,\tau_2 )\,\dDeltad (\tau_2 ,\tau_1) \nonumber\\
&
 \to& \int\int d^{d}x_1\, d^{d}x_2
 \, \frac{1}{2}(Z-1)_1 \,\frac{1}{2}(Z-1)_2\,
2\, {_\mu}\Delta_{\nu} (x_1,x_2) \,
  {_\nu}\Delta_{\mu} (x_2,x_1) \nonumber\\
&
 =& \int\int d^{d}x_1\, d^{d}x_2
 \, \frac{1}{2}(Z-1)_1 \,\frac{1}{2}(Z-1)_2\,
 2\, \Delta_{\mu \mu} (x_2,x_1) \,
  \Delta_{\nu \nu} (x_1,x_2)\,,
\label{wbn@c4*}\end{eqnarray}
the last line following from partial integrations.
For brevity, we have abbreviated
$[1-Z(\tau_i )]$ by
$(1-Z)_i$.
Using the field equation (\ref{wbn@n5})
and going back to one dimension
yields
\begin{equation}
 \langle {\cal A}^2_{\inter} \rangle_c
 = \frac{1}{2} \int\int d\tau_1 \, d\tau_2 \, (1-Z)_1\,
 (1-Z)_2\, \delta ^2 (\tau_1 ,\tau_2 ).
\label{wbn@c5}\end{equation}
To third order we calculate
\begin{eqnarray}
&&\!\!\!\!\!\!\!\! \langle {\cal A}^3_{\inter} \rangle_c
\! =\! \int\!\!\int\!\!\int\!\! d\tau_1\, d\tau_2 \, d\tau_3\,
 \frac{1}{2}(Z\!-\!1)_1 \,\frac{1}{2}(Z\!-\!1)_2\,
 \frac{1}{2}(Z\!-\!1)_3 \,8\,\dDeltad (\tau_1 ,\tau_2 )\,
 \dDeltad (\tau_2, \tau_3)\,\dDeltad (\tau_3, \tau_1)
\nonumber\\
&&
 \!\!\!\!\!\!\!\to \int\!\!\int\!\!\int\!\! d^{d}x_1\, d^{d}x_2 \,d^{d}x_3\,
 \frac{1}{2}(Z\!-\!1)_1 \,\frac{1}{2}(Z\!-\!1)_2\,
 \frac{1}{2}(Z\!-\!1)_3 \,8 \,
  {_\mu}\Delta_{\nu} (x_1,x_2) \,
  {_\nu}\Delta_{\sigma} (x_2,x_3)\,
  {_\sigma}\Delta_{\mu} (x_3,x_1)\nonumber\\
 &&
\!\!\!\!\!\!\!= -\int\!\!\int\!\!\int\!\! d^{d}x_1\, d^{d}x_2\,d^{d}x_3
 \, \frac{1}{2}(Z\!-\!1)_1 \,\frac{1}{2}(Z\!-\!1)_2\,
\frac{1}{2}(Z\!-\!1)_3\,
 8\, \Delta_{\mu \mu} (x_3,x_1) \,
  \Delta_{\nu \nu} (x_1,x_2)\,\Delta_{\sigma \sigma}(x_2,x_3) .
\nonumber\\
\label{wbn@c5*}\end{eqnarray}
 Applying again the field equation (\ref{wbn@n5})
and going back to one dimension, this
reduces  to
\begin{equation}
\langle {\cal A}^3_{\inter} \rangle_c
  = - \int\int\int d\tau_1 \, d\tau_2 \, d\tau_3\,(1-Z)_1\,
 (1-Z)_2\, (1-Z)_3
 \delta (\tau_1 ,\tau_2 )\,\delta (\tau_2, \tau_3)\,\delta (\tau_3, \tau_1).
\label{wbn@c6}\end{equation}
Continuing to $n$-order
and substituting Eqs.~(\ref{wbn@c4}), (\ref{wbn@c5}), (\ref{wbn@c6}),
etc. into (\ref{wbn@c2}),
we obtain  in the exponent of Eq.~(\ref{wbn@c2})
as sum
\begin{equation}
 - \langle {\cal A}_{\inter} \rangle_c +
 \frac{1}{2} \langle {\cal A}^2_{\inter} \rangle_c -
    \frac{1}{3!} \langle {\cal A}^3_{\inter} \rangle_c +\,
    \dots \,
  = \frac{1}{2}\,\sum_1^\infty\,\frac{c_n}{n},
\label{wbn@c7}\end{equation}
with
\begin{equation}
 c_n = \int d\tau _1 \dots d\tau _n \,
 C (\tau _1, \tau _2)\,C (\tau _2, \tau _3) \,
      \dots \, C  (\tau _n, \tau _1)
\label{wbn@c8}\end{equation}
where
\begin{equation}
 C (\tau, \tau') = [1-Z(\tau)]\,\delta (\tau, \tau').
\label{wbn@c9}\end{equation}
Substituting
this into Eq.~(\ref{wbn@c8})
and using the rule (\ref{wbn@4*})
yields
\begin{equation}
 c_n = \int \int d\tau _1 d\tau _n  \,[1-Z(\tau _1)]^{n}
 \,
    \delta ^2 (\tau _1 - \tau _n)
=  \delta (0) \int d\tau  \,[1-Z (\tau)]^n .
\label{wbn@c10}\end{equation}
Inserting these numbers into the expansion (\ref{wbn@c7}),
we obtain
\begin{eqnarray}
\!\!\!\!\!\!\!\!
 - \langle {\cal A}_{\inter} \rangle_c +
 \frac{1}{2} \langle {\cal A}^2_{\inter} \rangle_c -
    \frac{1}{3!} \langle {\cal A}^3_{\inter} \rangle_c +\,
    \dots \, % \nonumber\\&&
 & =& \frac{1}{2}\,\delta (0)\int d\tau\,
  \sum_1^\infty\,\frac{[1-Z (\tau)]^n}{n}
\nonumber \\& =&
 -  \frac{1}{2}\,\delta (0)\int d\tau\, \log Z (\tau),
\label{wbn@c11}\end{eqnarray}
which compensates precisely
the Jacobian factor $J$
in (\ref{wbn@c2}).

%%%%%%%%%%%%%%%

\end{document}